\documentclass[12pt,preprint]{aastex}

\slugcomment{Not to appear in Nonlearned J., 45.}

\shorttitle{Seyfert 1.5s Sample: IR dominated Eigenvector 1}
\shortauthors{Wang et al.}

\begin{document}


\title{A Sample of IRAS Infrared-Selected Seyfert 1.5 Galaxies: \\
      Infrared-Color $\alpha(60,25)$ Dominated Eigenvector 1}


\author{J. Wang and  J. Y. Wei}
\affil{National Astronomical Observatories, Chinese Academy of Sciences, Beijing, China
      }
\email{wj@bao.ac.cn}
\and
\author{X. T. He}
\affil{Department of Astronomy, Beijing Normal University, Beijing, China}



\begin{abstract}

The well-documented E1 relationships are first extended to infrared 
color $\alpha(60,25)$ and flux ratio 
[OIII]/H$\beta_{\rm{n}}$ by comparing emission line properties to
continuum properties in infrared wavelength.
Both direct correlations and a principle component 
analysis are used in a sample of 50 IRAS IR-selected Seyfert 1.5 galaxies.
In addition to confirm the correlations of E1 in Boroson \& Green (1992), 
our Eigenvector 1 turns out to be dominated by mid-infrared color $\alpha$(60,25), 
and most strongly effected by RFe, [OIII]/H$\beta_{\rm{n}}$, and EW(H$\beta_{\rm{b}}$).
Our analysis indicate that 
the objects with large E1 tend to co-existent with relatively young nuclear stellar populations, 
which implies that the E1 is related with nuclear starformation 
history. The IR-dominated Eigenvector 1 can be, therefore, inferred to be interpreted as 
``age'' of AGN. In confirmation of Xu et al. (2003), it is clear that the 
extreme Seyfert galaxies with both large RFe and large [OIII]/H$\beta_{\rm{n}}$ are 
rare in our Universe. 

\end{abstract}



\keywords{Galaxies: Seyfert -- Quasars: emission lines -- infrared: galaxies}


\section{Introduction}

   In order to understand the basic properties underlying the observed AGN's spectra, a
   number of consistent correlations among the observational parameters have been explored. 
   Among these properties, the Eigenvector 1 (E1) introduced by Boroson \& Green 
   (1992, hereafter BG92) is a milestone.
    These authors examined the properties 
   of a sample of 87 bright Palomar-Green (PG) quasars and found from a principal 
   component analysis (PCA) that the E1 correlates with the strength of FeII and anti-correlates 
   with the strength of the [OIII] line emission. Up to date,
   the best E1 correlation space involves (1) RFe, defined as flux ratio
   between the FeII complex and H$\beta$, and (2)FWHM of H$\beta$ broad component,
   supplemented by (3) soft X-ray photon index, $\Gamma_{\rm{soft}}$
   (e.g. Wang et al. 1996; Sulentic et al. 2000a; Xu et al. 2003; Sulentic et al. 2002;
   Grupe 2004, Laor et al. 1997; Lawrence et al. 1997; Grupe et al. 1999;
   Vaughan et al. 2001; Sulentic et al. 2004).

   At present, a great deal of evidence suggests that the Eddington ratio, defined as $L/L_{\rm{Edd}}$,
   is the principle physical driver of E1 (e.g. Sulentic et al. 2000a,b; Boroson 2002;
   Xu et al. 2003; Grupe 2004; Marziani et al. 2003b; BG92).
   Moreover, Marziani et al. (2001) reproduced an artificial
   relationship between RFe and FWHM of H$\beta$, the strongest anti-correlation in E1
   space, by combining some observational and numerical simulation results. In their
   process, the relation can be interpreted as driven by $L/L_{\rm{Edd}}$ associated
   with central black hole mass. The evolutionary significance of E1 was suggested
   that it might represent the ``age'' of AGN (Grupe et al. 1999; Grupe 2004).
   Narrow-Line Seyfert 1 galaxies (NLS1s) with large $L/L_{\rm{Edd}}$ and small black hole mass 
   might be young AGNs with more activity and in their earlier evolution phase (Mathur 2000).

   Over past few years there have been remarkable developments in understanding of connection between AGN
   and star formation in their host galaxies. A tight correlation between the mass of the black hole
   and the mass of galactic bugle within which it resides has been firmly established in past decade
   (e.g. Ferrarese \& Merritt 2000; Gebhardt et al. 2000; Marconi \& Hunt 2003; Magorrian et al. 1998; 
   Richston et al. 1998; Tremaine et al. 2002). 
   On the basis of this correlation many investigations indicated that there might be a close 
   connection between the formation of the black hole and that of its host galaxy 
   (Heckman et al. 2004 and references therein).

   If the black hole co-evolves with galactic bulge, and 
   If the E1 is related with the ``age'' of AGN as suggested by Grupe (2004), it is logically expected that 
   the nuclear stellar population is related with the E1 correlations. 
   This relationship could be directly 
   examined by investigating the nuclear stellar populations by optical stellar absorption 
   features (most notably the Balmer series and HeI lines) in broad line AGNs. 
   In fact, so far a few attempts to detect the nuclear stellar features in type I AGNs have been 
   made. Canalizo \& Stockton (2001) investigated the age of 
   nuclear stellar populations in
   the host of nine nearby QSOs by detection of Balmer absorption lines in the surrounding of the QSO or 
   nebular HII emission lines. The estimated ages range from a few Myr old to poststarburst ages.
   Brotherton et al. (1999) identified a stellar component with 400 Myr old instantaneous starburst
   in quasar UN\,J1025-0040. An exact Balmer jump was recently identified in NLS1 
   SDSS\,J022119.84+005628.4 (Wang et al. 2004).
   Zhou et al. (2005) recently compiled a sample of 74 poststarburst type I AGNs from
   Sloan Digital Sky Survey (SDSS). All these studies attempting to detect nuclear stellar population 
   have been limited to a few particular AGNs or poststarburst galaxies (approximately 1 Gyr old starbursts)
   with very deep Balmer absorptions and prominent Balmer jump. 
   However, a systematic spectroscopic study of circumnuclear stellar population of a large type I AGN sample 
   have been damped by the difficulty in 
   degenerating AGN and its hots galaxy. Because 
   both broad emission lines and AGN continuum mask the stellar features at optical band, 
   the detection of absorption lines of
   young stars at optical wavelength is usually quit difficult in most of Seyfert 1 galaxies or quasars.

   The question of the nuclear stellar population can be approached in an alternative way.
   The far-infrared warmness as quantified by the $\alpha$(60,25) index {\footnote {The infrared color index
   $\alpha(\lambda_1,\lambda_2)$ is defined as 
   $\alpha(\lambda_1,\lambda_2)=-\frac{\log[F(\lambda_2)/F(\lambda_1)]}{\log(\lambda_2/\lambda_1)}$, 
   where the wavelengths are in units of $\mu$m.}}
   has been usually
   used as an important tool to discriminate AGN and starburst activity in the nuclear or circumnuclear regions 
   of galaxies (de Grijp et al. 1985; Lipari 1994; Barthel 2001). Because of additional heating input from the 
   hard continuum radiation of central active nucleus, AGNs have warmer dust the do starburst regions (e.g. 
   Low et al. 1988). 
   The color $\alpha$(60,25), therefore, allows us to address
   the relative importance of AGN-activity and starburst-activity in not too distant active galaxies (see 
   section 4.1 for details).
   In this paper, we will use the infrared colors adopted from the IRAS survey to extend the well-known E1 
   relationships to AGN's infrared properties.

   There is another goal of this paper. Recently, a new relationship between flux ratio of [OIII] to 
   H$\beta$ narrow component and flux ratio of optical FeII blends to H$\beta$ broad component was found 
   in an AGN sample with high X-ray to optical flux ratio
   from ROSAT All-Sky Survey (RASS) (Xu et al. 2003).
   In their analysis, there were always
   large uncertainties in the determining the fraction of H$\beta$ emitting from the NLR.
   Deblending the optical permitted lines in a majority of cases is difficult because
   no transition between the narrow and broad components is observed. The anti-correlation
   between flux ratios $\rm{FeII/H\beta_{b}}$ and $[\rm{OIII}]/\rm{H\beta_{n}}$
   should be, therefore, investigated again and verified by involving a sample of Seyfert 1.5 galaxy (Osterbrock 1989, 
   Winkler 1992). Intermediate-type Balmer emission profile allows us to accurately extract narrow H$\beta$ 
   component from the integrated profile of emission line.

   In this paper, we examine the properties of optical emission lines of 50 bright Seyfert 1.5 galaxies
   and discuss their connection
   with infrared colors. These nearby AGNs were selected from the Catalog of quasars and active nuclei: 10th
   edition (Veron-Cetty \& Veron 2001) and the Catalog of the the IRAS Survey.
   Both direct correlations and PCA method are used to analyze the observed properties.

   The paper is organized as follows. The observational information and detailed
   spectral measurements are given in \S2. The statistical properties are investigated in \S3. Several
   diagnostically important correlations are studied. We discuss the implications of
   the results in \S4. A brief summary of main conclusions of this paper is given in the final section.
   A cosmology with cosmological constant: $\Omega_{\rm{m}}=0.3$, $\Omega_{\Lambda}=0.7$ and $h_0=0.7$
   is adopted for luminosity calculations.

\section{Observations and Emission Lines Measurements}

   The sample of objects was compiled from the Catalog of quasars and active nuclei:
   10th edition (Veron-Cetty \& Veron 2001) and IRAS survey. Because of the constraint of the observatory site and 
   instrumental capability, the objects were selected by requiring the objects to be in the northern 
   sky ($\delta>-10\symbol{23}$) and to be optical bright ($m_{v}\leq16.0\ \rm{mag}$). 
   Specially, there are total 66 such objects listed in Veron-Cetty \& Veron (2001). 
   Seven of the objects were not observed because of the poor weather conditions.

   Total 59 high quality spectra were obtained by using NAOC 2.16m telescope in Xinglong
   observatory during several observing runs performed from November 2001 to October
   2004. The observations were carried out by OMR spectrograph, using a Tektronix
   $1024\times1024$ CCD as detector. Two sets of spectra, denoted as set A and B, were involved
   in final emission lines measurements. For each object, the observation was performed with either 
   set A or B.
   The set A spectra were taken
   with a grating of 300 $\rm{g\ mm^{-1}}$ and a slit of 2\arcsec oriented in south-north
   direction. This setup provided a spectral resolution $\sim 10-11\AA$ as measured from comparison
   spectra. The observations of set B spectra with a resolution of $\sim5-6\AA$ were
   performed with a grating of 600 $\rm{g\ mm^{-1}}$ and a slit of 2.5\arcsec\ oriented in
   south-north direction. The blazed wavelengths were 6000\AA\ and 5400\AA\ in set A and B spectra, 
   respectively. Generally, the setup in set A provided a wavelength coverage of 3800-8700\AA\ in
   observed frame.
   This attempt covered both H$\beta$ and H$\alpha$ region in a single exposure in almost all set A
   spectra, except objects CSO\,409, 3C\,351 and 3C\,48.0, because of their large redshifts.
   On the other hand, only H$\beta$ region, from 4200\AA\ to
   6600\AA\ in observed frame, was covered in set B spectra.
   Exposure time was generally between 600-3600 seconds depending on the brightness of the object.
   In a majority of the observations, each object was exposed successively twice. The two frames were
   combined prior to extraction to enhance the S/N ratio and eliminate the contamination of cosmic-ray
   easily. The wavelength calibration associated with each object was carried out by helium-neon-argon
   comparison arcs obtained at the position being nearly identical to that of particular object.
   Two or three KPNO standard stars (Massey et al. 1988) were observed per night both for performing flux
   calibration and for the removal of atmospherical absorption features. All the objects were observed as close
   to meridian as possible. 

   In our final spectral data set, we find that the spectra of five out of 59 objects are contaminated strongly 
   by spectra of their host galaxies. In addition, there were no obvious broad H$\beta$ component 
   (with no or very faint H$\alpha$) 
   in other four objects. These 9 objects are finally excluded from our spectral measurements and from finally statistical 
   analyses. The log of observations of remaining 50 objects is given in Table 1.

   The unprocessed frames were reduced in standard procedures using IRAF package.
   The CCD reductions included bias subtraction, flatfield correction, and cosmic-ray
   removal before sign extraction. One-dimensional sky-subtracted spectra were
   wavelength and flux calibrated. For each of the standard stars, the observed flux 
   at wavelength 5000\AA\ was compared to its absolute flux. This procedure roughly yields that the 
   uncertainty of the flux calibration is generally no larger than 20\%. 
   The Galactic extinctions were corrected by color excess, 
   parameter E(B-V) from NED, assuming an $R_{V}=3.1$ extinction law (Cardelli et al. 1989). The spectra were 
   transformed to rest frame, along with K-correction, according to the narrow peak of H$\beta$.

   \subsection{FeII subtraction}

    It was long time to be known that the contamination of the FeII complex is a complicating
    factor in measuring strength and profile of particular emission lines such as H$\beta$,
    [OIII]$\lambda\lambda$4959,5007 and HeII$\lambda$4686. In order to reliably measure line properties
    and to determine the strength of the FeII blends it is necessary to appropriately
    model the FeII complex. The generally accepted model is the empirical template of the
    FeII complex described in BG92. The FeII template is the optical FeII
    emission in I\,ZW1, a bright NLS1 being of strong and narrow permitted FeII emission
    (Phillips 1978; Oke \& Lauer 1979; Veron-Cetty et al. 2004).
    Briefly, the template is a two-dimensional function of line width (FWHM)
    and intensity of the FeII blends.

    The same method described in BG92 was adopted in our measurements. The whole template
    is broadened to the FWHM of the broad component of H$\beta$ line by convolving with a Gaussian profile 
    and scaled to match the line intensity.
    Usually, the contribution of the FeII complex around H$\beta$
    can be well removed in terms of this template. The best subtraction of the
    FeII complex is derived by searching in the parameter space and producing a slick continuum
    at blue of H$\beta$ and between 5100\AA\ and 5500\AA\ after the subtraction. 
    The flux of the FeII complex is measured between the rest wavelength 4434\AA\ and 4684\AA\ as in BG92.
    
    The schemes of
    the FeII subtraction in two typical cases 3C\,48.0 and II\,Zw1 are illustrated in Figure 1.
    In each panel, the observed spectrum and resulting FeII-subtracted spectrum are displayed
    by the upper and middle curves, respectively. Note that the observed spectra are offset upward
    arbitrarily for visibility. The best adopted FeII templates are shown by bottom curves.

    \subsection{Profile modelling and measurements of other lines}

   The  FeII-contamination subtracted spectra are used to measure the non-FeII properties.
   The first step in the processing of the spectra is to remove the continuum from each object. The
   continuum is fitted by a power-law to the regions which seemed to be uncontaminated by emission lines.
   The two selected strong emission-line free windows are generally from $\lambda$4500 to $\lambda$4600
   and between 5100\AA-5500\AA. There are two parameters in determining each continuum component, i.e.,
   the flux and the slope.
   
   In our study, the observed line profiles are reproduced by a single Gaussian component
   and/or by a set of several Gaussian components because of their simplicity and practicability
   (Xu et al. 2003; Evans 1988; Rodriguez-Ardila et al. 2000). The line profiles are modelled
   and measured by SPECFIT task (Kriss 1996) in IRAF package. The fitting is persisted
   until the minimum of $\chi^{2}$, the measurement of goodness of profiling, is achieved.
   In each spectrum, the line profiles are modelled as follows. 1) Each of the forbidden lines
   [OIII]$\lambda\lambda$4959,5007 are modelled by a single Gaussian component.
   The atomic relationships,
   $F_{5007}/F_{4959}\doteq3$ (Storey \& Zeippen 2001) and $\lambda_{4959}/\lambda_{5007}=0.9904$, are
   used to reduce the number of free parameters and to improve the reliability of fitting procedure.
   2) The profile modelling of H$\beta$ line is straightforward in the spectra of objects of intermediate type.
   Each H$\beta$ profile is modelled by a set of two Gaussian components: a narrow and
   a broad Gaussian components. The broad component of H$\beta$ is referred as H$\beta_{\rm{b}}$ and
   is considered as representative of the broad line region (BLR). The narrow component is referred
   as H$\beta_{\rm{n}}$. 
   This adopted procedure accords with the generally accepted unified model of AGN in which the
   narrow and broad H$\beta$ components are emitted from NLR and BLR, respectively.

   The distribution of ratio between FWHM(H$\beta_{\rm{n}}$) and FWHM([OIII]) (both with instrumental 
   resolution correction, see \S3.1 for details) is illustrated in Figure 3. 
   A statistical analysis indicates that this ratio has a mean value of 1.26 and a standard deviation of 
   0.57. It is noted that there are a few objects
   in which a distinct narrow component is invisible in our sample.
   The interpretation 
   of the profile de-composition should be careful, although the observed profile can be fitted with 
   two Gaussian component adequately. Detailed examination of the results suggests that the flux of narrow 
   H$\beta$ is generally overestimated because the FWHM(H$\beta_{\rm{n}}$) is much large than that of [OIII] 
   line (e.g. 5 objects with FWHM(H$\beta_{\rm{n}}$)/FWHM([OIII])$>$2.0). 
   For these cases, we set the H$\beta_{\rm{n}}$ is unavailable and set the total H$\beta$ flux as 
   H$\beta_{\rm{b}}$ (see also in Wang et al. 2005, Zheng et al. 2002). 
   As described in BG92, the contribution of narrow component is rarely more than 3\% of total 
   H$\beta$ flux, even when there is a distinct narrow H$\beta$ emission.

   The modelling described above works well in most objects except in a few cases, for example the objects 3C\,390.3 and Mark\,509.
   The object 3C\,390.3 is a radio-loud (RL) Seyfert galaxy that shows double-peaked structure
   in their Balmer emission lines (e.g. O'Brien et al. 1998;  Eracleous \& Halpern 2003;
   Strateva et al. 2003). To measure the broad H$\beta$ emission of 3C\,390.3, we employ
   two separated broad Gaussian components to reproduce the observed double-peaked profile.
   In 3C\,390.3, the flux of H$\beta$ broad component includes the contributions of both broad
   Gaussian profiles. In Mark\,509, a very broad H$\beta$ component ($\rm{FWHM\sim10000\ km\ s^{-1}}$)
   must be acquired to reproduce the observed profile.
   Generally, the very broad emission line region (VBLR, $\rm{FWHM>7000\ km\ s^{-1}}$) is suggested to
   be optically thin to Balmer ionizing
   continuum and located at the inner edge of classical BLR (e.g. Corbin 1997a, b; Brotherton \& Wills 1994; 
   Kuraszkiewicz et al. 2004). 
   Sulentic et al. (2000c) reported a demise of classical BLR in luminous quasar
   PG\,1416-129. There remained a redward asymmetric very broad emission line component. These authors 
   suggested that the wing of H$\beta$ line are dominated by emission from optically thin VBLR.
   Moreover, the existence of VBLR is also
   supported by the observation of very broad and boxy profile of high ionized emission lines such
   as HeII$\lambda$4686 (Marziani et al. 2003a).
   
   As an illustration, the profile decomposition is shown in Figure 2 for two
   typical cases, MS\,0412.4-0802 and IRAS\,05078+1626. The observed spectra are represented by light solid lines,
   the modelled spectra by heavy solid lines. The narrow and broad components are shown by short and long
   dashed lines respectively. The lower
   panel underneath each spectrum illustrates the residual between the observed and modelled profiles.





\section{Results and Analysis}

   \subsection{Results and correlations}

   In order to investigate the statistical properties of a set of data, we list in Table 2
   the following items measured from spectra.
   Column (2) lists the equivalent width of $\rm{H\beta_{n}}$, and Column (3), the equivalent width of $\rm{H\beta_{b}}$.
   The equivalent width of the optical FeII complex measured between rest wavelength
   $\lambda4434$ and $\lambda4634$ is listed in Column (4). All equivalent width measurements refer to the continuum level
   at position of H$\beta$, $\lambda4861$. The continuum level is determined 
   from the modelled power-law continuum. Column (5) gives the FWHM,
   in units of $\rm{km\ s^{-1}}$ of $\rm{H\beta_{n}}$. Column (6) lists the FWHM of $\rm{H\beta_{b}}$.
   The FWHM of [OIII]$\lambda5007$ is listed in column (7).
   All quoted widths of emission lines were corrected for instrumental resolution.
   The intrinsic line widths were obtained by 
   $\Delta\lambda_{\mathrm{ture}}^2=\Delta\lambda_{\mathrm{obs}}^2-\Delta\lambda_{\mathrm{inst}}^2$
   there $\Delta\lambda_{\mathrm{inst}}$ is 
   the instrumental resolution that is determined from the width of night-sky emission lines and tested by 
   the spectra of comparison lamp. The next column lists the flux ratio $\rm{[OIII]/H\beta_{n}}$.
   The flux ratio, denoted as RFe, is listed in
   column (9). The RFe is defined as the flux ratio between the FeII complex and $\rm{H\beta_{b}}$. The velocity shift
   of central wavelength of $\rm{H\beta_{b}}$ with respect to that of $\rm{H\beta_{n}}$ is listed in column (10). Column (11)
   lists the velocity shift of Gaussian peak of [OIII] relative to that of $\rm{H\beta_{n}}$. In both column (10) and 
   (11), a positive velocity indicates a redshift of H$\beta_{\rm{b}}$ and a redshift of [OIII] line, respectively, 
   whereas a blueshift is inferred by a negative velocity.
   The column (12) and (13) list
   the far-infrared color $\alpha$(60, 100) and mid-far-infrared color $\alpha$(60, 25), respectively. 
   The luminosity of [OIII] emission line is listed in the last column.

   Having measured and tabulated a number of emission-line and continuum properties for this infrared selected Seyfert 1.5
   galaxies sample, we proceed to determine which features are correlated. This is approached by performing a correlation
   analysis on the particular parameters that are measured from lines and continuum.
   The Spearman rank-order correlation matrix, along with its significance matrix for measured properties, is calculated.
   The complete correlation coefficient matrix is presented in Table 3. The probability of the null correlation, $\rm{P_{s}}$,
   for a sample with corresponding correlation coefficient $\rm{r_{s}}$ is also shown in Table 3 for entries with
   $\rm{P_{s}<0.05}$. Each correlation coefficient is computed using only the objects for which both values are derived.  
   The coefficients involving RFe are calculated by survival analysis, because
   the RFe with zero value is non-detectable and treated as a lower limit in statistical analysis.

   \subsubsection{RFe vs. FWHM(H$\beta_{\rm{b}}$) and RFe vs.$L$([OIII])}

   The well-documented anti-correlation between RFe and FWHM(H$\beta$) has been confirmed in many studied AGN 
   samples (e.g. BG92; Xu et al. 2003; Grupe 2004; Sulentic et al. 2000, 2004). It is clear that our analysis reproduces 
   this anti-correlation which is moderately strong ($r_s=-0.321,P_s=0.0279$) in our moderately large sample of 
   type 1.5 Seyfert galaxy. This correlation is shown in the lower panel of Figure 4. BG92 found that their E1 has 
   significant correlation with RFe and [OIII] absolute magnitude. The similar anti-correlation 
   between strength of FeII and [OIII] 
   strength has also been detected by many authors (e.g. Zheng et al. 2002; Vaughan et al. 2001).
   In the present data, we detected a relatively strong 
   ($r_s=-0.386, P_s=0.0081$) anti-correlation between RFe and [OIII] luminosity. The upper panel of Figure 4 illustrates 
   this correlation.

    \subsubsection{Correlations with IR color index $\alpha$(60,25)}

   In addition to confirm the E1 correlations in BG92, newly discovered correlations involving the IR color 
   index $\alpha(60,25)$ are found in our sample. These correlations are displayed in Figure 5.
   Figure 5(a) displays the anti-correlation between the $\alpha$(60,25) and RFe.
   The anti-correlation is strong ($r_{s}=-0.518, P_{s}=0.0003$) in the present data whereas it is only
   weak ($r_{s}=-0.140$) if the color is compared directly with EW(FeII). This means that the correlation
   between $\alpha$(60,25) and RFe arises because EW(H$\beta_{\rm{b}}$) gets larger as the mid-infrared warmth gets hotter.
   The correlation between $\alpha$(60,25) and [OIII]/H$\beta_{\rm{n}}$ is plotted in Figure 5(b).
   The calculated correlation coefficient is $r_{s}=-0.538$, corresponding to a significance level
   $P_{s}=0.0012$. This plot shows a trend for the objects with higher ratio
   [OIII]/H$\beta_{\rm{n}}$ to have hotter mid-infrared warmth. The similar correlation is found by Keel et al. (1994)
   in a large sample of ``warm'' Seyfert 2 galaxy.
   The correlation between $\alpha$(60,25) and EW(H$\beta_{\rm{b}}$) is illustrated in Figure 5(c).
   A Spearman rank-order test yields a correlation coefficient $r_{s}=0.590$ ($P_{s}<10^{4}$).
   In Figure 5(c), the lower the value of EW(H$\beta_{\rm{b}}$), the cooler the mid-far-infrared color.
   The strong ($r_s=0.478, P_s=0.0009$) correlation between $\alpha(60,25)$ and $L$([OIII]) is illustrated 
   in Figure 5(d). When the mid-infrared warmth increases, the luminosity of [OIII] line increases.

  \subsection{Principle component analysis}

   In fact, some significant relations are due to selection effects and the fact that
   we have chosen to measure essentially the same properties in somewhat different ways (see e.g. BG92; Xu et al. 2003;
   Grupe 2004). The correlation between EW(FeII) and RFe ($r=0.692, P_s<10^{-4}$) are due to dependent parameters.
   In addition, the correlations between different measures of same property (e.g. EW(H$\beta_{\rm{b}}$) and RFe) and 
   another property (e.g. $\alpha(60, 25)$) are degenerated.
   To investigate the correlations more meaningfully, we perform a PCA
   on the correlation matrix. The PCA is applied to 36 out of 49 objects using the following eight parameters: 
   the EW(H$\beta_{\rm{b}}$), parameter RFe, FWHM of H$\beta_{\rm{b}}$, flux ratio [OIII]/H$\beta_{\rm{n}}$,
   $\rm{\Delta V(H\beta)}$, $\rm{\Delta V([OIII])}$, infrared colors $\alpha$(60, 100) and $\alpha$(60,25), and L[OIII]
   each of which contains potentially unique information. The Spearman (rank-order) correlation matrix is used to 
   run PCA program. Rank values are used rather than the measured properties, because the influence of extreme objects
   can be minimized by this nonparameter analysis (Brotherton 1996).

   Table 4 shows the results of PCA.
   The first four components together account for marginally more than
   70\% of the variance, and the first principle component seems to be an important underlying parameter that 
   governs the observed properties of the AGN. The PCA indicates that the first principle component is dominated by
   mid-infrared color $\alpha(60,25)$. It has a projection -0.841 on the Eigenvector 1 while the next high 
   projection is only 0.224 on Eigenvector 4. 
   In addition, the Eigenvector 1 has a strong effect on RFe, [OIII]/H$\beta_{\rm{n}}$, and EW(H$\beta_{\rm{b}}$). 
   Other three properties, FWHM(H$\beta_{\rm{b}}$), $\Delta \rm{V([OIII])}$ and $L$([OIII]), are also correlated 
   with Eigenvector 1.

   \subsection{Other properties} 

   Figure 6(a) displays the anti-correlation between the $\Delta \rm{V([OIII])}$ and FWHM(H$\beta_{\rm{b}}$) 
   ($r_{s}=-0.343, P_{s}=0.0230$). There is a trend that objects with wider H$\beta$ line tend to have larger velocity 
   shift of [OIII] with respect to the H$\beta$ narrow component. The correlation between the $\Delta \rm{V([OIII])}$ 
   and RFe is shown in Figure 6(b) ($r_{s}=-0.320, P_{s}=0.0319$).

   Xu et al. (2003) reported a significant ($r_s=0.600,P_s<10^4$) anti-correlation between the RFe and 
   flux ratio [OIII]/H$\beta_{\rm{n}}$ by
   investigating an AGN sample with high X-ray-to-optical flux ratio from RASS. Our statistical analysis indicates that
   this correlation is weak in our sample, however.  The lack of strong correlation in our
   sample is most likely due to under-sampling.
   Figure 7 is plot of the RFe vs. flux ratio [OIII]/H$\beta_{\rm{n}}$ for our sample ($r_{s}=-0.252$) and sample in 
   Xu et al. (2003). The data presented in this paper are marked with solid squares, and Xu et al. (2003) sample with 
   open triangles. 
   
   Although the correlation in our sample is not as strong as that obtained by Xu et al. (2003), we find that the point 
   distribution of our sample is similar to that shown by Xu et al. (2003). The digram is
   not evenly populated for both two samples. It is clear that almost all the objects in two samples are 
   predominately found in the region being under a solid line drawn by eye, and 
   the extreme region with large RFe and large flux ratio of 
   [OIII] to H$\beta_{\rm{n}}$ is restricted.

   The distribution of the line flux ratio [OIII]/H$\beta_{\rm{n}}$ of the objects in our sample is displayed in
   Figure 8.
   We find a mean value of 7.02 and a standard variance of 3.58. Generally, the characteristic ionization state
   of gas photoionized by the Seyfert 1 nucleus is of very high [OIII]/H$\beta_{\rm{n}}\sim 10$. However,
   very low ionized NLR can be identified in a fraction of the objects in our sample. For 36 objects with measured 
   flux ratio [OIII]/H$\beta_{\rm{n}}$ in our sample, 
   there are 5 objects (i.e. 1H\,1934-063; IRAS\,00580+3055; MCG\,+08-11-011; MCG\,+08-15-056;
   NMP1G\,05.0216) with [OIII]/H$\beta_{\rm{n}}<3$. It indicates that these 5 objects
   can not entirely classified as Seyfert 2 galaxies when their BLR are obscured by the torus on the line-of-sight
   (Veilleux \& Osterbrock 1987). As discussed below, there are two most likely interpretations of these 
   extremely small ratio of [OIII]/H$\beta_{\rm{n}}$: [OIII] suppression due to high density as suggested 
   by Xu et al. (2003) and contamination of starburst.





\section{Discussion}

\subsection{Connection between Eigenvector 1 and nuclear starformation history}

   It is noted that an eigenvector is always specific to a certain sample depending on which observed
   parameters have been used and the range of parameters. There is, therefore, no one ``Eigenvector 1''
   for every AGN sample.
   In our infrared-selected Seyfert 1.5 galaxies sample, the Eigenvector 1 is dominated by mid-infrared color 
   $\alpha(60,25)$. The most prominent correlations indicates the trends for objects with ``cold'' mid-infrared
   color to have small flux ratio [OIII]/H$\beta_{\rm{n}}$, small EW(H$\beta_{\rm{b}}$),
   low luminosity of [OIII], and large value of RFe. 

   The understanding of role of the infrared color $\alpha(60,25)$ is not straightforward.
   Generally, the infrared color $\alpha(60,25)$ allows us to determine the relative importance of AGN-activity and
   starburst-activity in a galaxy. Low et al. (1988) pointed out that in many cases AGNs display
   a warmer dust than do starforming regions. This warmer dust is heated by an additional heating caused by
   an intense radiation emitted from the nucleus, and strongly influence the mid-infrared flux at 25$\mu$m.
   De Grijp et al. (1985) reported that the infrared color $\alpha$(60,25) can be used as
   an efficient parameter to discriminate AGN and starburst activity in the nuclear and circumnuclear
   regions of galaxies. A high fraction of Seyferts among galaxies can be obtained with $\alpha(60,25)>$-1.5.
   Kewley et al. (2001) assessed the relative importance of AGN- and starburst-activity in type II AGN by combining
   the infrared color with line ratio [OIII]/H$\beta_{\rm{n}}$. They found that the AGNs have
   on average a warmer infrared color than starbursts. Hao et al.(2005) indicated that the infrared selected 
   QSOs with significant far-IR excess contributed by starburst have smaller color index $\alpha(60, 25)$. 
   As additional evidence, 
   Gonzalez Delgado et al. (2001) found that, in Seyfert 2 galaxies, the nuclei with young and intermediate-age
   population show lower ionization level (smaller flux ratio [OIII]/H$\beta_{\rm{n}}$) and cooler $\alpha$(60,25)
   than the nuclei dominated by an old stellar population.
   When the relative importance of starburst increases, our results indicate that the FeII/H$\beta_{\rm{b}}$ 
   ratio increases, and EW(H$\beta_{\rm{b}}$), [OIII]/H$\beta_{\rm{n}}$ and $L$([OIII]) decrease.

   The far-infrared color-color diagram has been generally used as an efficient tool to discriminate 
   between different types of activity in the nuclear and circumnuclear starforming region of galaxies. 
   As an additional test, the IRAS infrared color-color diagram is shown in Figure 9. In the same diagram,
   the power-law line is also indicated. The AGN and starburst loci are shown by short and long dashed lines,
   respectively. The AGN locus is adopted from Canalizo \& Stockton (2001), and the starburst locus from 
   Lipari (1994, and references therein). The size of each point is proportional to the projection on 
   Eigenvector 1 of the corresponding object. It is noted from Figure 9 that the objects with small projection 
   on Eigenvector 1 are mainly located in QSO/Seyfert region, while the objects showing large projection   
   are distributed in/near the starburst region.

   Some comments on the objects located in/near the starburst region are discussed more detailed as follows: 
  
   \begin{itemize}
   \item \sl 3C\,48.0,\hspace{0.3cm} \rm This object displays 
   morphological evidence for a recent merger event with a likely double nuclei and a tidal tail extending
   several arcseconds (Canalizo \& Stockton 2000, Stockton \& Ridgway 1991, Chatzichristou et al. 1999). 
   Very young stellar populations were suggested in the 
   second nucleus about 1\symbol{125} northeast of the QSO (Canalizo \& Stockton 2000). Recent near infrared observations
   of the second nucleus 
   indicated the presence of warm dust that is likely heated by starbursts or ongoing starformation 
   (Zuther et al. 2004).

   \item \sl NGC\,3227,\hspace{0.3cm} \rm This object is a nearby Seyfert galaxy interacting with a gas-poor companion. 
   A clear 3.3$\mu$m polycyclic aromatic hydrocarbon (PAH) emission feature, which 
   is usually associated with starformation activity, was identified in its inner 60pc
   ($\mathrm{EW_{3.3}=52\pm10\AA}$, Rodriguez-Ardila \& Viegas 2003). 
   These authors also observed another two objects: Mark\,766 and 
   1H\,1934-063 that are listed in our sample. In Mark\,766, the PAH feature has been detected as well 
   ($\mathrm{EW_{3.3}=40\pm11\AA}$), while no evidence for PAH feature has been observed in 1H\,1934-063
   ($\mathrm{EW_{3.3}<10\AA}$). The position of 1H\,1934-063 in the color-color diagram is indicated by
   an open star. As a first inspection, both NGC\,3227 and Mark\,766 have much cooler $\alpha(60,25)$ than 
   does 1H\,1934-063.   

   \item \sl NGC\,7469,\hspace{0.3cm} \rm This object is an luminous infrared source. 
   Its circumnuclear (1\symbol{125}.5-2\symbol{125}.5) ring 
   structure has been observed at radio, optical,  near-infrared, and mid-infrared bands (see references in 
   Davies, Tacconi \& Genzel 2004). Using the 2.3$\mu$m CO 2-0 absorption and continuum slope, 
   Davies, Tacconi \& Genzel (2004) resolved the nuclear star cluster (0\symbol{125}.15-0\symbol{125}.20) 
   and found that the age of the cluster is less than about 60 Myr.   

   \item \sl Mark\,732,\hspace{0.3cm} \rm Cid Fernandes et al. (1998) obtained the spatial resolved 
   long slit spectra of this object. 
   Circumnuclear young stellar features are identified in the spectrum 
   of region being north-east of the nucleus (4\symbol{125} away from the nucleus).

   \end{itemize}

   In summary, in our IR-selected AGN sample, the Eigenvector 1 not only holds the correlations 
   appeared in BG92's E1, but also first
   extends the well-documented E1 space into infrared color $\alpha$(60, 25) and ratio [OIII]/H$\beta_{\rm{n}}$. 
   This extension suggests that the E1 relates with nuclear 
   starformation history, and that the E1 can be interpreted as ``age'' of AGN as suggested by Grupe (2004).
   When the Eigenvector 1 increases, AGN with larger RFe tends to co-existent with relatively young 
   stellar populations.

   Both observational evidence and theoretical scenario indicate that starburst is a key piece of the AGN machinery
   through evolutionary processes. The understanding of co-existence of AGN and star formation in their host
   galaxies is accumulating. Powerful circumnuclear starbursts ($<$Gyr) have been
   unambiguously identified in $\sim$40\% of nearby Seyfert 2 galxies by several optical
   spectroscopic investigations of moderately large samples of type 2 Seyfert nuclei (e.g.
   Joguet et al.2001; Gonzalez Delgado, Heckman \& Leitherer 2001; Cid Fernandes et al. 2001,
   Storchi-Bergman et al. 2001). Gonzalez Delgado (2002) summarized the evidence in favor of
   the co-existence of AGN and starburst activity by showing the observations at optical and
   near-infrared band. Theoretical studies also provided a plausible picture about the co-evolution 
   of AGN and circumnuclear starburst (e.g. Granato et al. 2004). 
   In the early phase of AGN development, the star-formation rate is high,
   which means there are plenty of gases that can be attracted into the nucleus of galaxy under
   the gravitation of the central black hole. These gases are ultimately
   accreted onto black hole and lead to a high accretion rate. At latter phase, 
   the accretion rate decreases because of the diminishing of content of gas.

   PCA is a quantitative method for replacing a group of variables with a new variable. This group of 
   variables might be generally measuring the same driving principle. 

   Generally, the ratio [OIII]/H$\beta_{\rm{n}}$ (and also $L$([OIII])) is orientation independent
   because both lines are emitted only from optically thin region at large enough distance from
   the nucleus to be unobscured by the material causing the anisotropy in the featureless nuclear continuum
   and in the broad lines (Jackson et al. 1989). The isotropy of [OIII] line emission has
   been questioned by some studies of radio-loud AGNs (Baker 1997; di Serego Alighieri et al. 1997).
   However, the correlation between [OIII] and orientation independent [OII] emission
   indicates that the [OIII] emission is not dependent on orientation effect (Kuraszkiewicz 2000).
   The mid-far-infrared color is also a nominally unbiased isotropic property. It
   is a reasonable first approximation, although this may not be strictly true if the torus
   is optically thick in the mid-infrared (e.g. Pier \& Krolik 1992; Granato \& Danese 1994).
   On the basis of these results, it is not likely that the correlations in our Eigenvector 1
   is only due to the orientation effect.

   At present, the E1 is mostly 
   interpreted as governed by $L/L_{\rm{Edd}}$ (e.g. Boroson 2002; Grupe 2004). It is, therefor, reasonable 
   to expect that our Eigenvector 1 is most possibly driven by $L/L_{\rm{Edd}}$ as well.
   The parameter RFe is usually considered as a good tracer  
   of $L/L_{\rm{Edd}}$ (BG92; Boroson 2002; Marziani et al. 2001; Zamanov \& Marziani 2002, Netzer et al. 2004)
   and scales with $L/L_{\rm{Edd}}$ as $\rm{RFe}\propto 0.55\log(L/L_{\rm{Edd}})$ (Marziani et al. 2001).
   If so, the anti-correlation between RFe and mid-far-infrared color $\alpha$(60,25) possibly reveals 
   a linkage between $L/L_{\rm{Edd}}$ and nuclear starformation history, and that 
   $L/L_{\rm{Edd}}$ is large in AGN with relatively strong starburst-activity (e.g. Hao et al. 2005).   
   In order to test this scenario, $L/L_{\rm{Edd}}$ and black hole mass are necessary to be 
   calculated. This calculation is, however, not convincing in our Seyfert 1.5 sample, because of the influence 
   of orientation effect which obscures a large fraction of continuum emitting from the central engine. 
   The role of $L/L_{\rm{Edd}}$ will be tested in the next work by a larger sample of type I AGN . 

   On the basis of above discussions, we further conjecture that the correlation between RFe and $\alpha$(60,25)
   implies an evolutionary trace of Seyfert galaxies and QSOs. Cid Fernandes et al. (2001) speculated that 
   the composite starburst plus Seyfert 2 nucleus system evolve into pure Seyfert 2 galaxies.
   As analogue with the evolutionary scenario of Ultra infrared luminous galaxies (ULIGs)
   suggested by Sanders et al. (1988; see also in Hao et al. 2005 and references therein), we suspect 
   that the ``cold'' Seyfert galaxies possibly evolve into ``warm'' Seyfert galaxies.

   \subsection{RFe-[OIII]/H$\beta_{\rm{n}}$ space}

   As shown in \S3.2, both our sample and the sample in Xu et al. (2003) show that the Seyfert galaxies are 
   not uniformly populated in the digram plotted as [OIII]/H$\beta_{\rm{n}}$ vs. RFe. 
   We note that there are significant number of objects with very small [OIII]/H$\beta_{\rm{n}}$ for the 
   sample of Xu et al. (2003). This phenomenon, however, does not appear in our sample.  
   These very low [OIII]/H$\beta_{\rm{n}}$ might be caused by the overestimate of H$\beta_{\rm{n}}$ component as 
   described in \S1.  
   Xu et al. (2003) speculated that the anti-correlation between [OIII]/H$\beta_{\rm{n}}$ 
   and RFe is mainly driven by density effect.  
   Although this explanation can not be proved or excluded by our analysis in current data set, 
   we propose that 
   the contamination of starforming can also be a possible interpretation.
   In fact, as shown in \S3.1, both [OIII]/H$\beta_{\rm{n}}$ and RFe are tightly correlated 
   with $\alpha(60,25)$. When RFe is large, the increased relative importance of starburst could 
   reduce the observed value of [OIII]/H$\beta_{\rm{n}}$.

  \subsection{Outflows: a clue of linkage between BLR and NLR}

   The detection of [OIII] velocity shift is difficult to be doubted for following two reasons: 1) in a majority of cases,
   the resulting wavelength accuracy is better than 1\AA. The corresponding uncertainty of velocity is less than
   $100\ \rm{km\ s^{-1}}$; 2) the narrow components of H$\beta$ can be modelled accurately because of the apparent profile
   inflection.

   Zamanov et al. (2002) found that the large [OIII] blueshift relative to H$\beta_{\rm{n}}$
   ($\Delta \rm{V}([OIII])>250\ \rm{km\ s^{-1}}$) is confined to sources with
   $\rm{FWHM(H\beta_{b})}\leq 4000\ \rm{km\ s^{-1}}$. These authors interpreted the large blueshift 
   of [OIII] with respect to H$\beta_{\rm{n}}$ as a result of outflow whose origin is 
   considered to be strong radiation pressure due to large $L/L_{\mathrm{Edd}}$.
   Their purely kinematical model suggested that
   the [OIII]$\lambda$5007 blueshifts are associated with the high-ionization outflow generated in objects with
   high accretion rate. Many theoretical studies suggested that a high $L/L_{\rm{Edd}}$ is prefer to result in 
   strong mass outflow. 
   King \& Pounds (2003) proposed a model in which 
   winds in quasars are probably produced when the black holes accrete at or above the Eddington limit.
   The dynamical model of accretion disk wind
   indicates that the vertical outflow ratio strongly depends on the accretion rate.
   The magnitude of outflow increases
   significantly with the accretion rate, assuming a constant viscosity parameter (Witt, Czerny \& Zycki 1997).
   The simplest, and mostly possible, interpretation of the correlation between [OIII] shift amplitude $\Delta V(\rm{[OIII]})$
   and RFe is, therefore, that we are clear seeing a dynamical linkage between the outflow of emission material and central 
   accretion power in terms of a relatively strong wind arising from the central region of AGN.

\section{Conclusions}

   This paper presents the statistical study of optical emission-line properties of a sample of 50 Seyfert 1.5 galaxies.
   The line properties are compared with their continuum properties in infrared wavelength to extend the documented 
   E1 relationships into infrared property of AGN. The statistical
   analysis allows us to draw following conclusions:

   \begin{enumerate}

      \item We first extend the E1 relationships into infrared color $\alpha(60,25)$ and flux ratio
      [OIII]/H$\beta_{\rm{n}}$. 
      In addition to confirm the BG92's E1, the PCA analysis indicates that our Eigenvector 1 is 
      dominated by mid-far-infrared color $\alpha$(60,25).
      The color $\alpha$(60,25) is found to be strongly anti-correlated with RFe
      ($r_{s}=-0.518; P_{s}=0.0003$), and correlated with EW(H$\beta$) ($r_{s}=-0.590; P_{s}<10^{-4}$),
      [OIII]/H$\beta_{\rm{n}}$ ($r_{s}=0.538; P_{s}=0.0012$),
      and $L$([OIII]) ($r_{s}=0.478; P_{s}=0.0009$). The Eigenvector 1 is inferred to increase with 
      relative importance of starburst with respect to AGN activity, and to be, therefore, related with nuclear
      starformation history.
      We further suspect that a ``cold'' Seyfert galaxy perhaps evolves into a ``warm'' one.

      \item Although the relationship between RFe and [OIII]/H$\beta_{\rm{n}}$ in our sample is not as
      significant as that obtained by Xu et al. (2003), the populations of points are same for both our sample 
      and Xu et al (2003). It is clear that almost all the objects are located below an envelope. This effect 
      implies that Seyfert galaxies rarely have both large value of RFe and large ratio of [OIII]/H$\beta_{\rm{n}}$.

      \item Other two correlations,  FWHM(H$\beta_{\rm{b}}$) vs. $\rm{\Delta V[OIII]}$ and RFe vs. $\rm{\Delta V[OIII]}$, 
      are identified in our sample. According to the simple kinematic model, these two correlations could 
      provide a direct linkage between BLR and NLR.  

      \item In our sample [OIII]/H$\beta_{\rm{n}}<$3 are identified in 5 out of 36 objects with determined narrow component 
      of H$\beta$.  These five objects can not
      be classified as Type 2 AGNs when their BLR are obscured on the line-of-sight.

   \end{enumerate}

\acknowledgements

   We are grateful to Profs. Todd A. Boroson and Richard F. Green for providing us the FeII template.
   We thank an anonymous referee for many useful suggestion and his careful and critical review, 
   D. W. Xu, C. N. Hao, Y. M. Mao for valuable discussion and help. Special thanks go to the staff at
   Xinglong observatory as a part of National Astronomical Observatories, China Science Academy
   for their instrumental and observational help. This research has made use of the NASA/IPAC
   Extragalactic Database (NED), which is operated by the Jet Propulsion Laboratory, Caltech, under
   contact with the National Aeronautics and space Administration. This work was supported by NFS of 
   China (Grant number: 19990754, 10473013 and 10503005).

\begin{figure}
   \centering
   \includegraphics[width=11cm]{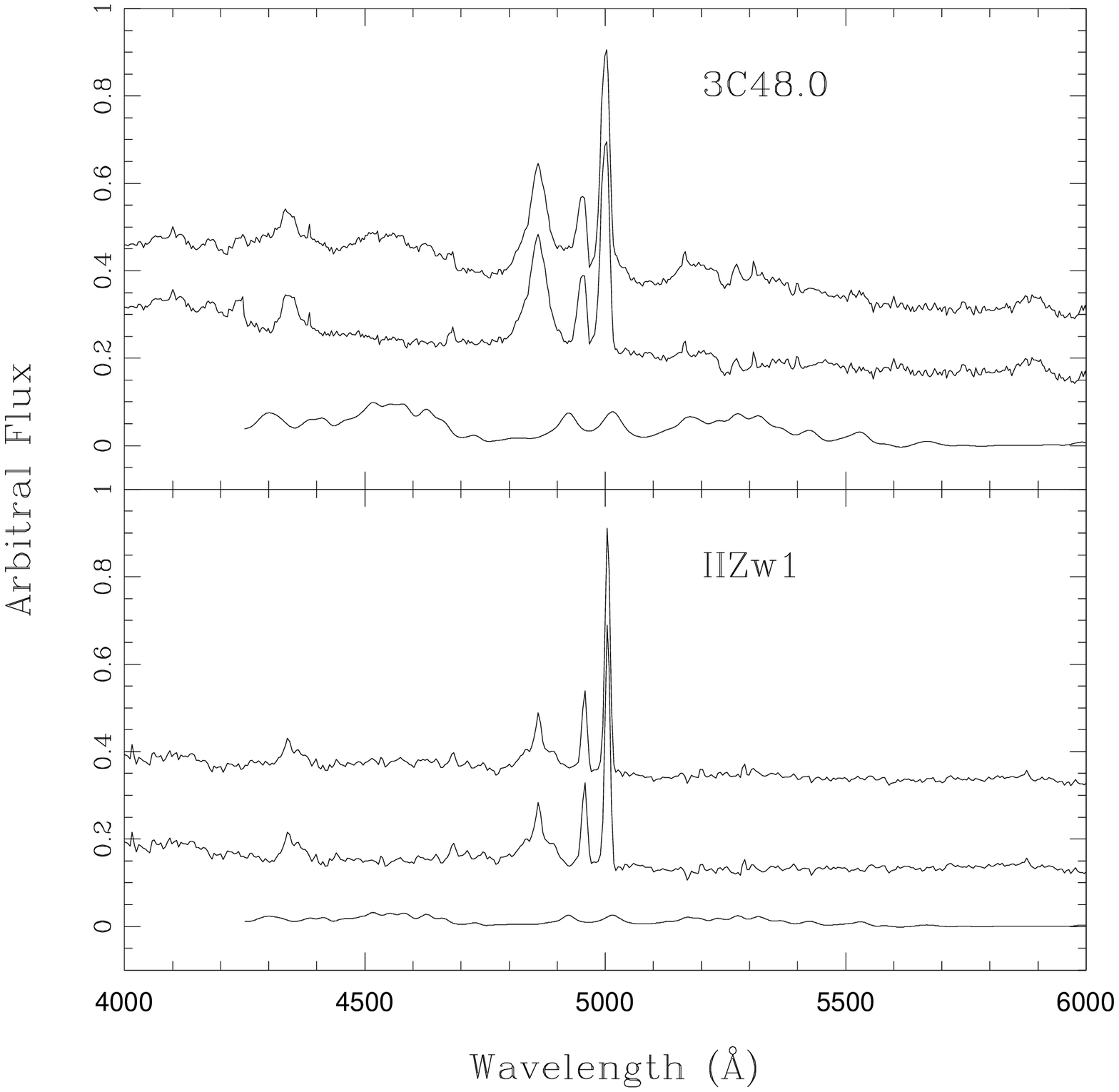}
   \caption{An illustration of the FeII subtraction in two typical cases, 3C\,48.0 and
    II\,Zw1. In each panel, the top curve is the observed spectrum which is shifted
    upward by an arbitrary amount for visibility. The middle curve is the FeII-subtracted
    spectrum, bottom curve the best adopted FeII template.}
\end{figure}

 \begin{figure}
   \centering
   \includegraphics[width=11cm]{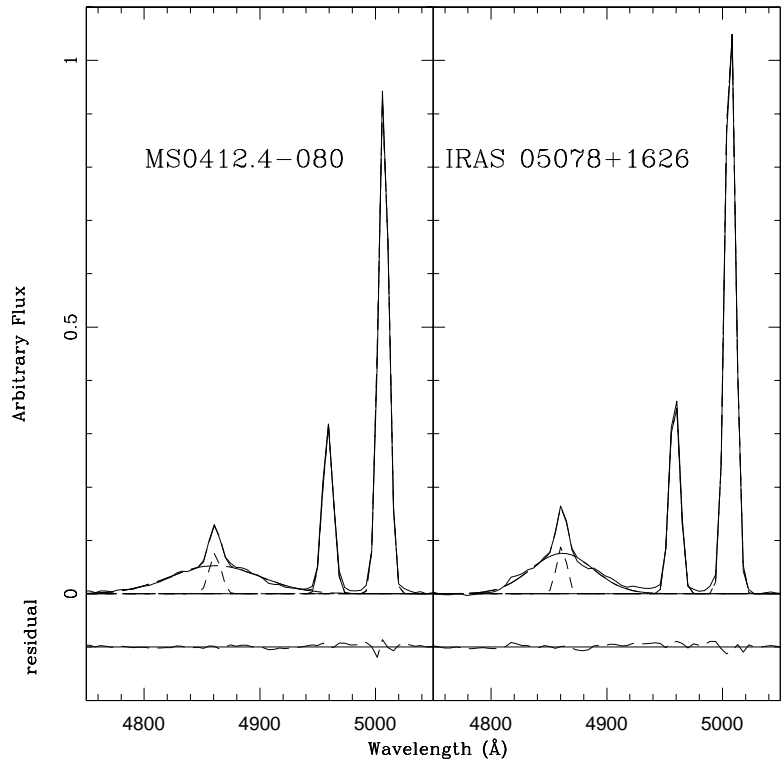}
   \caption{Profile modelling and decomposition for two typical objects, MS\,0412.4-0802 and IRAS\,05078+1626.
    The observed and modelled profiles are represented by light and heavy solid lines, respectively. The narrow peaks
    of H$\beta$ are displayed by short dashed lines for visibility, the broad H$\beta$ components by long dashed
    lines. The lower sub-panel underneath each spectrum
    illustrates the residuals between the observed and fitted emission profile.}
   \end{figure}

   \begin{figure}
   \centering
   \includegraphics[width=11cm]{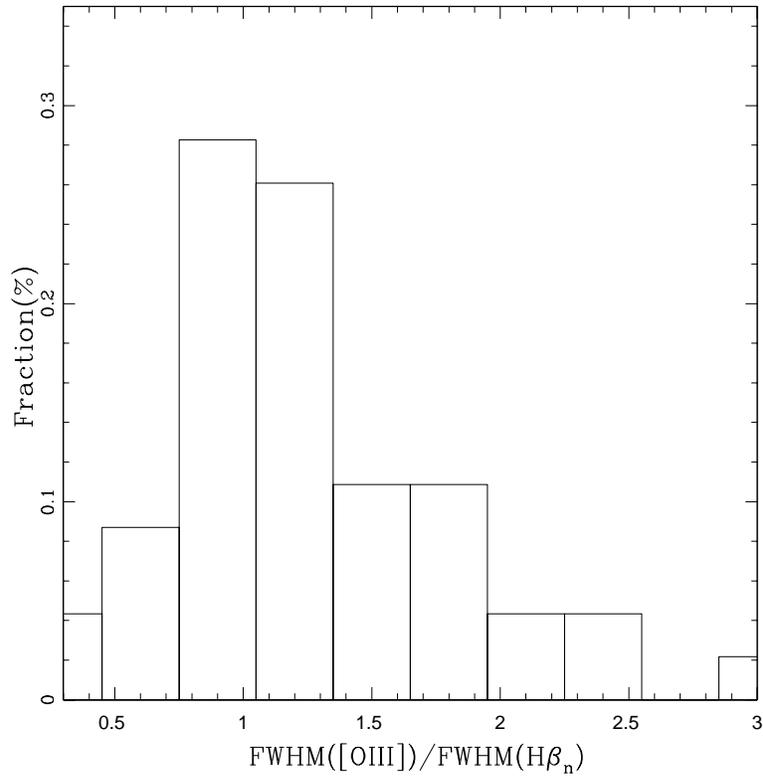}
   \caption{The fraction of the ratio of FWHM(H$\beta_{\rm}$) and FWHM([OIII]).} 
   \end{figure}

   \begin{figure}
   \centering
   \includegraphics[width=11cm]{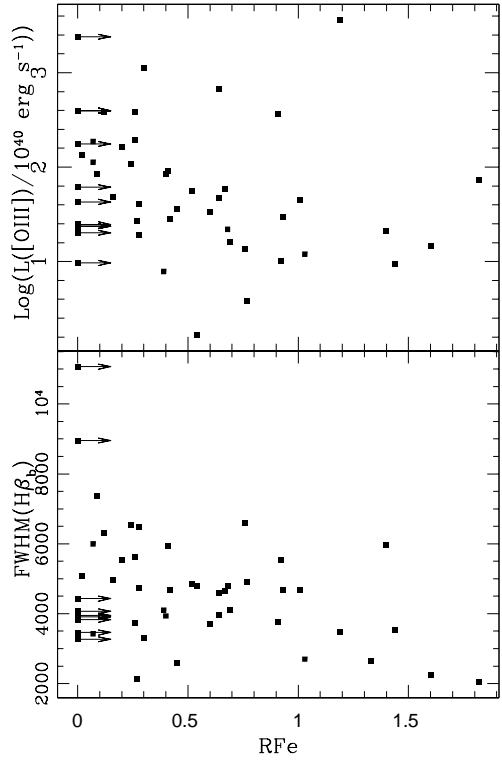}
   \caption{
   \sl Lower panel: \rm RFe plotted against the FWHM of H$\beta_{\rm{b}}$ for all the object in our sample 
   ($r_s=-0.321, P_s=0.0279$). \sl Upper panel: \rm Plot of RFe vs. $L$([OIII]) ($r_s=-0.386, P_s=0.0081$). In both panels, 
   the points with zero fluxes of FeII are indicated by superposed arrows.
   }
   \end{figure}

   \begin{figure}
   \centering
   \includegraphics[width=15cm]{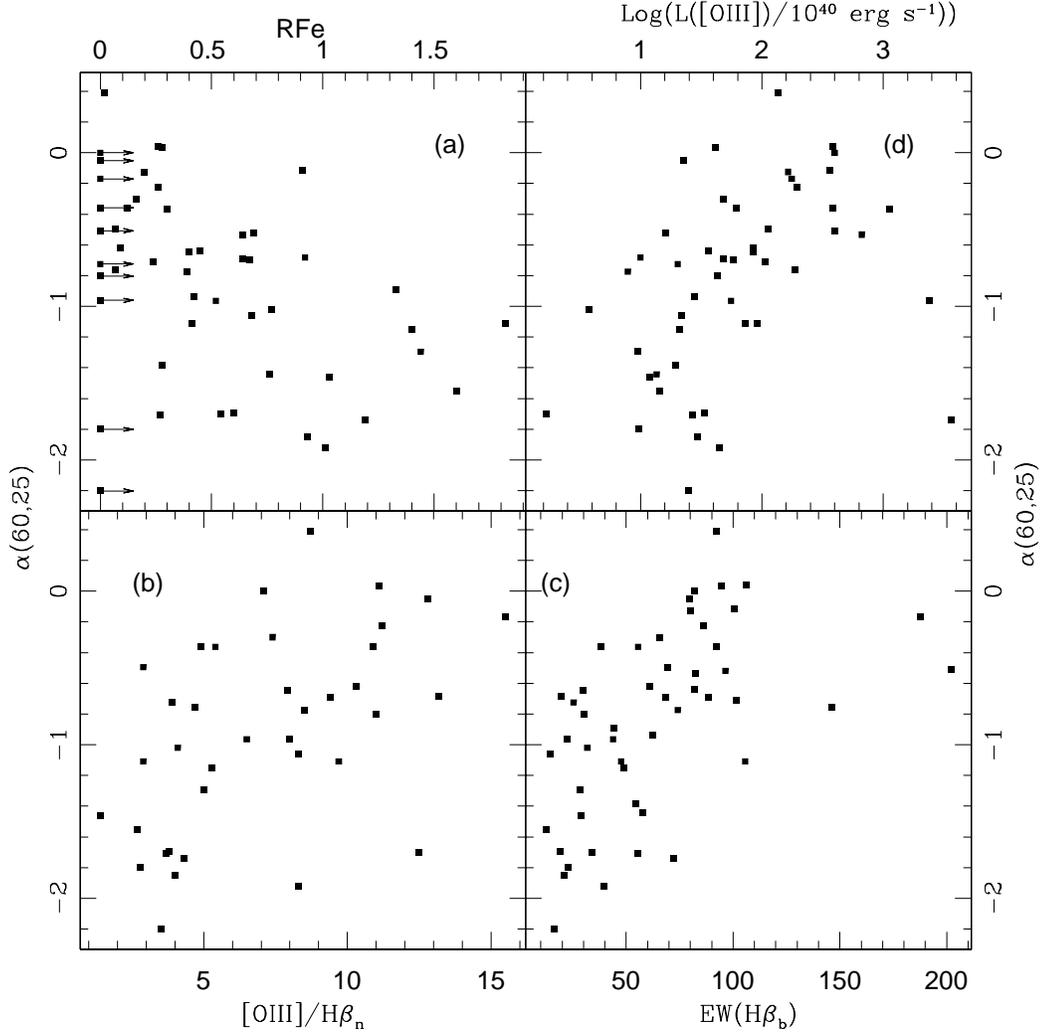}
   \caption{(a): Plot of mid-infrared color $\alpha$(60,25) vs. RFe 
   ($r_{s}$=-0.518,$P_{s}=0.0003$). The correlation coefficient is calculated
   by survival analysis, because the data with zero FeII flux is nondetectable and considered to be lower limit of
   its true value. The points with zero fluxes of FeII are indicated by superposed arrows. 
   (b): Mid-infrared color $\alpha$(60,25) plotted against flux ratio [OIII]/H$\beta_{\rm{n}}$ 
   ($r_{s}$=0.538,$P_{s}=0.0012$). (c): Mid-infrared color $\alpha$(60,25)
   plotted against equivalent width of H$\beta$ broad component($r_{s}$=0.590,$P_{s}<10^{-4}$). 
   (d): $L$([OIII]) plotted against $\alpha$(60,25) ($r_s=0.478, P_s=0.0009$).
   }
   \end{figure}

   \begin{figure}
   \centering
   \includegraphics[width=16cm]{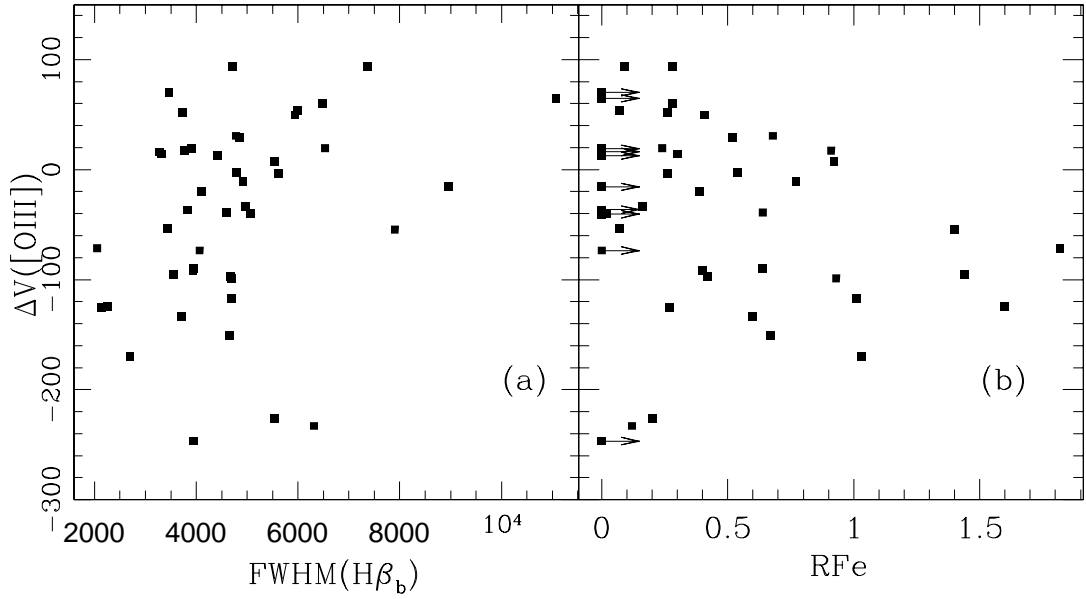}
   \caption{
   (a) FWHM(H$\beta_{\rm{b}}$) plotted against the velocity shift of [OIII] with respect to the peak
   of H$\beta$ narrow component, $\Delta V([\rm{OIII}])$ ($r_{s}$=-0.343, $P_{s}=0.0230$). (b): The RFe plotted 
   against $\Delta V([\rm{OIII}])$ ($r_{s}$=-0.320, $P_{s}=0.0319$). 
   The survival analysis is carried out to take place the ordinary Spearman rank-order analysis.
   The points considered to be lower limits of RFe are indicated by arrows.
   }
   \end{figure}

   \begin{figure}
   \centering
   \includegraphics[width=11cm]{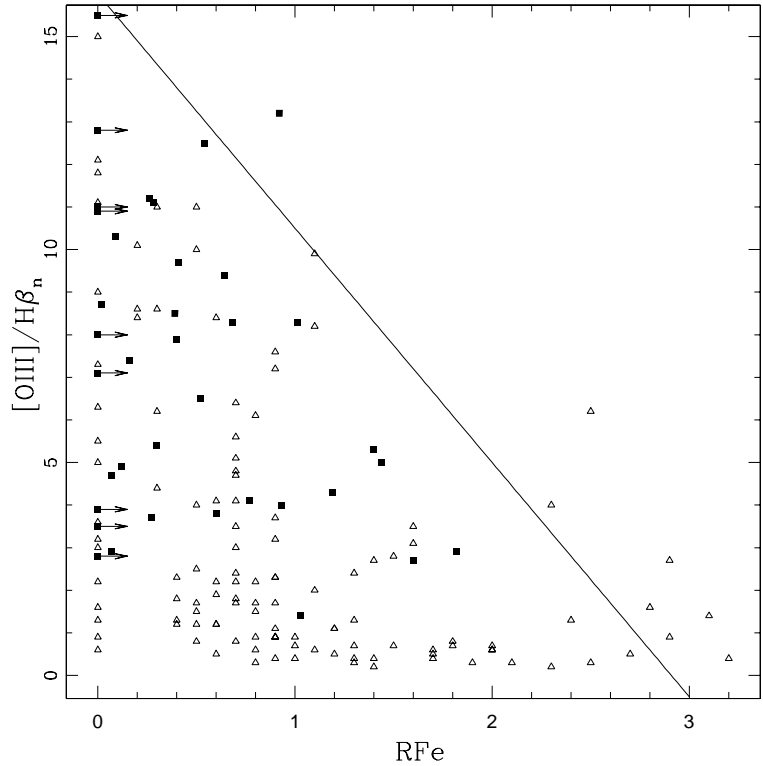}
   \caption{
    RFe plotted against the flux ratio [OIII]/H$\beta_{\rm{n}}$ ($r_{s}$=-0.252) for 36 out of 50 objects in our sample and 
   for data in Xu et al. (2003). Our sample is denoted by solid square, and sample of Xu et al. (2003) by open triangle. 
   The superposed arrows indicate that the corresponding values of
   the RFe is the lower limits. The solid line from left-top to right-bottom indicates the position of the envelope drawn by eye.
   }
   \end{figure}

   \clearpage

   \begin{figure}
   \centering
   \includegraphics[width=11cm]{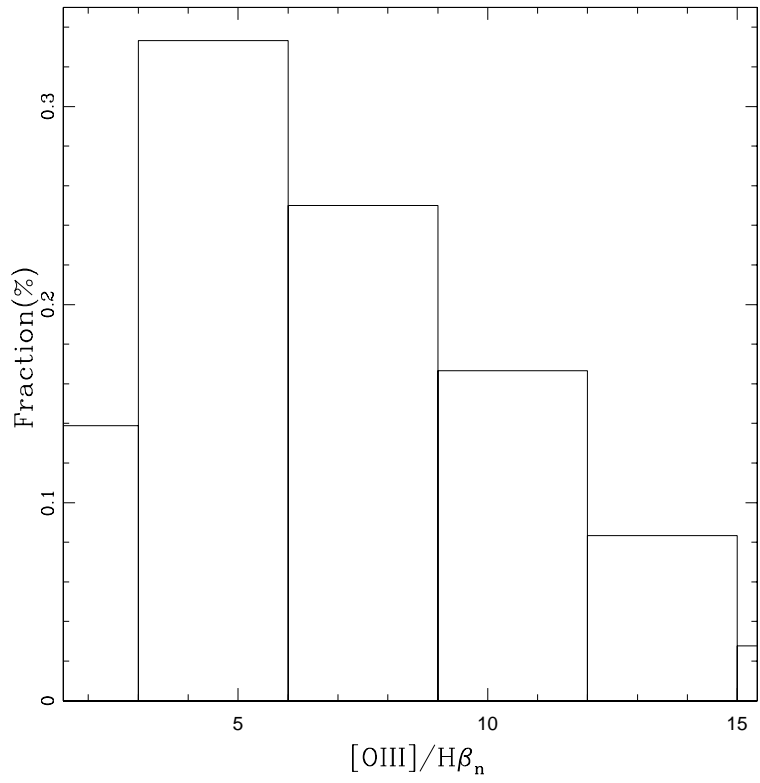}
   \caption{Distribution of line flux ratio [OIII]/H$\beta_{\rm{n}}$ for 36 objects with determined H$\beta$ narrow component in 
   our sample.
   }
   \end{figure}

   \begin{figure}
   \centering
   \includegraphics[width=11cm]{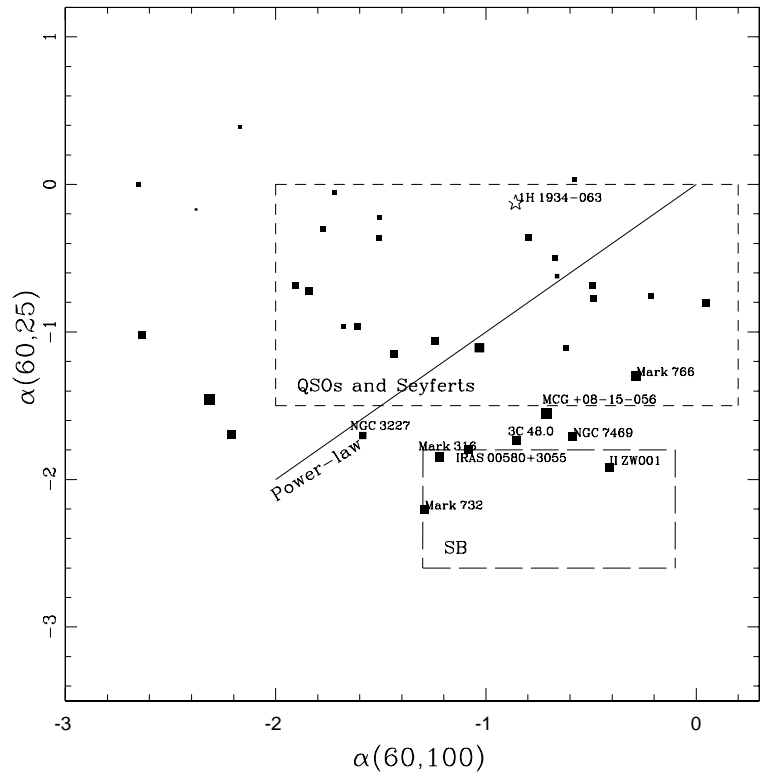}
   \caption{IRAS color-color diagram. The localizations of AGN and starburst regions are displayed by 
    short and long dashed lines, respectively. The size of each point is proportional to the projection 
    on Eigenvector 1 of that object. 
   }
   \end{figure}






\clearpage



\clearpage
   \begin{deluxetable}{lccccccccc}
   \rotate
   \tablewidth{0pt}
     \tabletypesize{\scriptsize}
   \tablecaption{Log of Spectroscopic Observation}
   \tablehead{
   \colhead{Object name}  & \colhead{Set}     &  \colhead{R.A.} &  \colhead{Dec} & 
   \colhead{z}            & \colhead{Date}  &  \colhead{Exp}  &  \colhead{$\rm{m_{v}}$} &
   \colhead{$\rm{M_{B}}$} & \colhead{AGN Type}\\
   \colhead{(1)} & \colhead{(2)} & \colhead{(3)} & \colhead{(4)} & 
   \colhead{(5)} & \colhead{(6)} & \colhead{(7)} & \colhead{(8)} & 
   \colhead{(9)} & \colhead{(10)}
   }
   \startdata
   Mark\,543        & B & 000226.4 & +032106 & 0.025518 & 2004 Sep 17 & 1200 & 14.68 & -20.6 & S1.5\\
   IRAS\,00580+3055 & A & 010048.0 & +311200 & 0.051100 & 2001 Nov 19 & 1800 & 15.80 & -21.6 & S1.5\\
   II\,Zw001        & A & 012159.8 & -010224 & 0.054341 & 2001 Nov 17 & 1500 & 15.17 & -21.8 & S1.5\\
   3C\,48.0         & A & 013741.3 & +330935 & 0.367000 & 2001 Nov 19 & 2100 & 16.20 & -25.2 & QSO S1.5\\
   NGC\,985         & A & 023437.8 & -084715 & 0.043143 & 2001 Oct 12 & 1800 & 14.28 & -22.4 & S1.5\\
   NGC\,1019        & A & 023827.4 & +015428 & 0.024187 & 2001 Nov 17 & 2000 & 14.95 & -20.1 & S1.5\\
   NGC\,1275        & A & 031948.2 & +413042 & 0.017559 & 2001 Nov 17 &  900 & 12.48 & -21.9 & S1.5\\
   MS\,0412.4-0802  & A & 041452.7 & -075540 & 0.037910 & 2001 Oct 12 & 1500 & 14.91 & -21.1 & S1.5\\
   3C\,120          & A & 043311.1 & +052116 & 0.033010 & 2001 Nov 17 & 1600 & 15.05 & -20.8 & S1.5\\
   NPM1G\,-05.0216  & A & 044720.7 & -050814 & 0.044200 & 2001 Nov 19 & 1500 & 15.11 & -21.2 & S1.5\\
   IRAS\,05078+1626 & A & 051045.5 & +162956 & 0.017880 & 2001 Nov 17 & 2000 & 15.64 & -19.4 & S1.5\\
   MCG\,+08-11-011  & A & 055453.6 & +462621 & 0.020484 & 2001 Nov 16 & 1200 & 14.62 & -20.0 & S1.5\\
   Mark\,6          & A & 065212.3 & +742538 & 0.018813 & 2001 Nov 16 & 1200 & 14.19 & -20.1 & S1.5\\
   Mark\,376        & A & 071415.1 & +454156 & 0.055980 & 2001 Nov 18 & 1650 & 14.62 & -22.5 & S1.5\\
   IRAS\,F07144+4410& A & 071800.6 & +440527 & 0.061440 & 2001 Nov 16 & 1500 & 15.50 & -223. & S1.5\\
   Mark\,9          & A & 073657.0 & +584613 & 0.039874 & 2001 Nov 18 & 1000 & 14.37 & -22.1 & S1.5\\
   B3\,0754+394     & A & 075800.1 & +392029 & 0.096000 & 2001 Dec 17 & 1800 & 14.36 & -24.1 & QSO S1.5\\
   MCG\,+08-15-056  & A & 081516.8 & +460430 & 0.040975 & 2001 Nov 19 & 1500 & 15.20 & -21.7 & S1.5\\
   NGC\,3227        & A & 102330.6 & +195156 & 0.003860 & 2001 Nov 18 &  600 & 11.79 & -18.7 & S1.5\\
   NGC\,3516        & A & 110647.4 & +723407 & 0.008840 & 2001 Nov 19 &  500 & 12.40 & -20.5 & S1.5\\
   Mark\,732        & A & 111349.8 & +093510 & 0.029230 & 2004 May 21 & 1050 & 14.17 & -21.4 & S1.5\\
   Mark\,1447       & A & 113029.1 & +493458 & 0.095900 & 2002 Feb 12 & 2400 & 16.00 & -22.8 & S1.5\\
   Mark\,766        & A & 121826.7 & +294847 & 0.012930 & 2002 Feb 11 & 1200 & 13.57 & -20.0 & S1.5\\
   Mark\,1320       & A & 121908.8 & -014829 & 0.103000 & 2002 Feb 13 & 1800 & 15.00 & -24.0 & QSO S1.5 \\
   PG\,1351+640     & A & 135315.7 & +634546 & 0.088200 & 2004 May 22 & 2700 & 14.28 & -24.1 & QSO S1.5\\
   Mark\,662        & A & 135406.4 & +232549 & 0.055000 & 2002 Feb 11 & 1800 & 15.37 & -21.6 & S1.5\\
   CSO\,409         & A & 140438.8 & +432707 & 0.323300 & 2002 Feb 15 & 2000 & 15.62 & -25.9 & QSO S1.5\\
   OQ\,208          & A & 140700.4 & +282715 & 0.076576 & 2002 Feb 12 & 1650 & 15.35 & -22.2 & S1.5\\      
   NGC\,5548        & A & 141759.5 & +250812 & 0.017175 & 2004 May 21 & 1200 & 13.73 & -20.7 & S1.5\\ 
   H\,1419+480      & B & 142129.8 & +474725 & 0.072296 & 2003 Feb 02 & 3000 & 15.40 & -22.8 & S1.5\\
   CGCG\,163-074    & A & 143209.0 & +313505 & 0.055108 & 2002 Feb 13 & 1650 & 15.50 & -22.1 & S1.5\\
   Mark\,817        & A & 143622.1 & +584739 & 0.031455 & 2002 Feb 14 & 1200 & 13.79 & -22.3 & S1.5\\
   Mark\,841        & A & 150401.2 & +102616 & 0.036422 & 2002 Feb 11 & 1500 & 14.27 & -22.2 & S1.5\\
   SBS\,1527+564    & A & 152907.5 & +561605 & 0.099000 & 2002 Feb 14 & 2100 & 15.80 & -23.1 & QSO S1.5 \\
   Mark\,290        & A & 153552.3 & +575409 & 0.029577 & 2002 Feb 13 & 1800 & 14.96 & -20.7 & S1.5\\
   Mark\,871        & A & 160836.4 & +121951 & 0.033657 & 2002 Feb 15 & 2100 & 14.94 & -20.9 & S1.5\\
   TON\,0256        & A & 161413.2 & +260416 & 0.131000 & 2004 May 22 & 3000 & 15.41 & -23.5 & QSO S1.5 \\
   3C\,351          & A & 170441.5 & +604428 & 0.371940 & 2002 Feb 14 & 2200 & 15.28 & -26.5 & QSO S1.5 \\
   UGC\,10683B      & A & 170500.4 & -013230 & 0.032000 & 2004 May 23 & 2400 & 15.55 & -19.5 & S1.5\\
   RX\,J1715.9+3112 & A & 171602.0 & +311223 & 0.111000 & 2004 May 22 & 3600 & 16.00 & -23.2 & QSO S1.5\\
   IRAS\,17216+3633 & A & 172323.3 & +363010 & 0.040000 & 2004 May 21 & 3600 & 16.00 & -20.9 & S1.5\\
   3C\,390.3        & A & 184209.0 & +794617 & 0.056100 & 2003 Oct 25 & 3600 & 15.38 & -21.6 & S1.5\\
   1H\,1934-063     & A & 193733.2 & -061306 & 0.010587 & 2001 Sep 21 & 1200 & 15.35 & -17.8 & S1.5\\
   Mark\,509        & B & 204409.7 & -104324 & 0.034397 & 2004 Sep 17 & 2400 & 13.12 & -23.3 & QSO S1.5\\
   3C\,445.0        & A & 222349.7 & -020613 & 0.056200 & 2001 Nov 17 & 1500 & 15.77 & -20.8 & S1.5\\
   MR\,2251-178     & A & 225405.9 & -173455 & 0.063980 & 2001 Nov 19 &  900 & 14.36 & -23.1 & QSO S1.5\\ 
   NGC\,7469        & A & 230315.6 & +085226 & 0.016317 & 2001 Sep 21 &  600 & 13.04 & -21.6 & S1.5\\
   Mark\,315        & A & 230402.7 & +223727 & 0.038870 & 2001 Nov 17 & 1200 & 14.78 & -21.3 & S1.5\\
   Mark\,316        & A & 231340.5 & +140115 & 0.040902 & 2001 Nov 18 & 1200 & 15.20 & -21.7 & S1.5\\
   NGC\,7603        & A & 231856.6 & +001438 & 0.029524 & 2001 Nov 18 & 1000 & 14.01 & -21.5 & S1.5\\
   \enddata
   \tablecomments{Col.(1), Object name. Col.(2), Set of spectra described in the text. Col.(3), Object's
    right ascension. Col.(4), Declination of object. Col.(5), Redshift from NED. Col.(6), Date
    of observation. Col.(7), Averaged exposure time in seconds. Col.(8), apparent magnitude in
    V band. Col.(9), Absolute magnitude at B band from Veron-Cetty \& Veron(2001). Col.(10),
    Classification type of AGN}
   \end{deluxetable}

   \begin{deluxetable}{lccccccccccccc}
   \rotate
   \tablewidth{0pt}
   \tabletypesize{\scriptsize}
   \tablecaption{List of Emission Line Properties}
   \tablehead{
    \colhead{Name}             & \colhead{$\rm{H\beta_{n}}$} & \colhead{$\rm{H\beta_{b}}$}  &  \colhead{FeII} & \colhead{RFe}&
    \colhead{$\rm{H\beta_{n}}$} & \colhead{$\rm{H\beta_{b}}$}  & \colhead{[OIII]} & 
    \colhead{[OIII]/H$\beta_{\rm{n}}$}  &  \colhead{$\Delta \rm{V(H\beta_{\rm{b}})}$}\tablenotemark{b} & 
    \colhead{$\Delta \rm{V([OIII])}$}\tablenotemark{c} & \colhead{$\alpha(100,60)$} & \colhead{$\alpha(60,25)$} & 
     \colhead{$\rm{\log(\frac{L([OIII])}{10^{40}\ erg\ s^{-1}})}$}\\
        & EW(\AA) & EW(\AA) & EW(\AA) & & FWHM\tablenotemark{a} & FWHM\tablenotemark{a} & FWHM\tablenotemark{a} &    
     &($\rm{km\ s^{-1}}$)& ($\rm{km\ s^{-1}}$)     &      &    & \\
    \colhead{(1)} & \colhead{(2)} & \colhead{(3)} & \colhead{(4)} & \colhead{(5)} & 
    \colhead{(6)} & \colhead{(7)} & \colhead{(8)} & \colhead{(9)} & \colhead{(10)} & 
    \colhead{(11)}& \colhead{(12)}& \colhead{(13)} & \colhead{(14)}
   } 
   \startdata

    Mark\,543        &  3.6 &  25.4 &  0.0 &  0.00 &  264.0 &  3827.8 &  256.5 &  3.9 &  -25.9 &  -36.4 & -1.841 & -0.724 & 1.305\\ 
    IRAS\,00580+3055 &  9.0 &  22.7 &  0.0 &  0.00 &  656.7 &  3460.3 &  515.1 &  2.8 & -165.0 &   70.2 & -1.084 & -1.800 & 0.985\\ 
    II\,Zw001        &  6.3 &  39.6 & 39.8 &  1.01 &  499.3 &  4693.4 &  545.2 &  8.3 & -184.4 & -117.4 & -0.413 & -1.918 & 1.649\\
    3C\,48.0         & 11.2 &  72.1 & 85.6 &  1.19 & 1361.8 &  3473.2 & 1087.9 &  4.3 & -275.5 & -444.1 & -0.856 & -1.739 & 3.558\\
    NGC\,985         &  6.5 & 105.6 & 43.0 &  0.41 &  617.9 &  5933.2 &  597.5 &  9.7 &   -9.4 &   49.9 & -0.617 & -1.109 & 1.962\\ 
    NGC\,1019        &  5.1 &  31.8 & 24.6 &  0.77 &  584.8 &  4914.6 &  372.6 &  4.1 & -297.5 &  -11.0 & -2.637 & -1.019 & 0.579\\
    NGC\,1275        & 10.3 &  30.1 &  0.0 &  0.00 &  502.3 &  3950.2 & 1654.7 & 11.0 &  -62.5 & -246.7 &  0.046 & -0.803 & 1.632\\
    MS0412.4-0802    & 16.2 &  86.3 & 22.7 &  0.26 &  539.1 &  5618.8 &  406.1 & 11.2 &   17.9 &   -3.8 & -1.506 & -0.224 & 2.287\\
    3C\,120          &\dotfill&101.6& 23.9 &  0.24 & 1443.6 &  6530.0 &  686.2 &\dotfill&783.4 &   19.6 & -1.358 & -0.711 & 2.029\\
    NPM1G\,-05.0216  &  9.2 &  29.1 & 30.0 &  1.03 &  426.9 &  2699.0 &  693.2 &  1.4 &  169.1 & -169.6 & -2.317 & -1.458 & 1.077\\
    IRAS\,05078+1626 & 15.3 &  79.4 &  0.0 &  0.00 &  330.1 &  3904.1 &  308.4 & 12.8 &   26.5 &   19.1 & -1.718 & -0.052 & 1.357\\
    MCG\,+08-11-011  & 38.1 &  69.2 &  4.9 &  0.07 & 1416.3 &  6001.5 &  820.2 &  2.9 &  922.1 &   53.2 & -0.672 & -0.495 & 2.051\\
    Mark\,6          & 15.3 &  60.8 &  5.7 &  0.09 &  894.7 &  7374.8 & 1014.7 & 10.3 & -350.2 &   93.6 & -0.661 & -0.621 & 1.927\\
    Mark\,376        &\dotfill&96.5 & 67.0 &  0.69 &\dotfill&  4094.0 &  542.7 &\dotfill&\dotfill&\dotfill&-1.054& -0.520 & 1.209\\
    IRAS\,F07144+4410&\dotfill&106.2& 27.1 &  0.26 & 1450.7 &  3733.8 &  957.3 &\dotfill&-467.1&   52.1 & -2.881 &  0.039 & 2.581\\
    Mark\,9          &\dotfill& 81.8& 36.9 &  0.45 &\dotfill&  2582.5 &  715.9 &\dotfill&\dotfill&\dotfill&-0.688& -0.638 & 1.562\\
    B3\,0754+394     &\dotfill& 82.3& 52.8 &  0.64 & 1580.3 &  4599.7 &  644.6 &\dotfill&-165.9&  -38.8 & -2.781 & -0.533 & 2.824\\
    MCG\,+08-15-056  &  8.5 &  12.4 & 19.8 &  1.60 &  626.3 &  2247.0 &  571.3 &  2.7 &   64.2 & -124.0 & -0.712 & -1.553 & 1.160\\
    NGC\,3227        &  5.2 &  33.9 & 18.3 &  0.54 &  612.2 &  4778.9 &  746.4 & 12.5 & -332.5 &   -2.9 & -1.586 & -1.702 & 0.220\\
    NGC\,3516        &  2.7 &  74.2 & 29.1 &  0.39 &  634.5 &  4101.6 &  591.7 &  8.5 & -198.0 &  -19.7 & -0.491 & -0.773 & 0.896\\
    Mark\,732        &  3.8 &  16.2 &  0.0 &  0.00 &  872.7 &  4427.9 &  686.9 &  3.5 &  226.5 &   12.7 & -1.292 & -2.201 & 1.394\\
    Mark\,1447       &  2.4 &  48.8 & 68.5 &  1.40 &  645.0 &  7902.7 &  815.3 &  5.3 &-1145.3 &  -54.4 & -1.436 & -1.150 & 1.319\\
    Mark\,766        & 22.9 &  28.7 & 41.3 &  1.44 &  830.5 &  3544.9 &  428.2 &  5.0 & -302.2 &  -95.5 & -0.285 & -1.296 & 0.978\\
    Mark\,1320       &  6.9 &  30.0 & 12.0 &  0.40 &  565.8 &  3934.7 &  409.3 &  7.9 & -613.3 &  -91.8 & -3.474 & -0.646 & 1.929\\
    PG\,1351+640     &  7.2 &  38.4 &  4.4 &  0.12 &  707.8 &  6316.1 &  986.4 &  4.9 &  -18.3 & -233.0 & -0.799 & -0.358 & 2.581\\
    Mark\,662        &  1.0 &  19.3 & 17.7 &  0.92 &  417.8 &  5536.8 &  782.9 & 13.2 &   40.2 &    7.8 & -1.908 & -0.683 & 1.000\\
    CSO409\tablenotemark{d}&\dotfill&44.4 & 59.2 &  1.33 &\dotfill&  2641.0 &\dotfill&\dotfill&\dotfill&\dotfill&-0.918& -0.889&\dotfill\\
    OQ\,208          &  2.1 &  68.5 & 43.9 &  0.64 &  326.5 &  3951.1 &  764.1 &  9.4 & 1719.6 &  -90.1 & -0.493 & -0.687 & 1.679\\
    NGC\,5548        & 10.5 &  92.5 &  0.0 &  0.00 &  556.6 &\dotfill &  609.0 & 10.9 &\dotfill&  -40.5 & -0.997 & -0.358 & 1.789\\
    H\,1419+480      &\dotfill&80.1 & 16.4 &  0.20 & 1006.0 &  5531.4 &  541.4 &\dotfill&-239.6& -226.3 & -0.859 & -0.125 & 2.217\\
    CGCG\,163-074    &  3.3 &  14.6 & 10.0 &  0.68 &  878.4 &  4779.2 &  699.3 &  8.3 &  -24.7 &   30.7 & -1.241 & -1.062 & 1.340\\
    Mark\,817        &\dotfill&88.5 & 59.3 &  0.67 & 1618.0 &  4661.3 &  663.3 &\dotfill& 13.7 & -150.4 & -0.034 & -0.694 & 1.765\\
    Mark\,841        &  3.4 &  94.3 & 26.7 &  0.28 &  723.7 &  6485.2 &  773.6 & 11.1 & -145.4 &   59.7 & -0.580 &  0.033 & 1.613\\
    SBS\,1527+564    & 20.7 &  82.1 &  0.0 &  0.00 &  480.5 &  4064.7 &  586.5 &  7.1 &   69.0 &  -73.5 & -2.652 &  0.000 & 2.599\\
    Mark\,290        &  5.8 &  65.7 & 10.6 &  0.16 &  487.2 &  4972.4 &  482.3 &  7.4 &  -37.8 &  -33.6 & -1.773 & -0.300 & 1.682\\
    Mark\,871        &\dotfill&62.5 & 26.1 &  0.42 &  475.3 &  4670.4 &  328.5 &\dotfill&-58.8 &  -96.9 & -0.432 & -0.937 & 1.447\\
    TON\,0256        & 15.4 &  55.5 & 16.6 &  0.30 &  798.2 &  3307.6 &  479.3 &  5.4 &   43.3 &   14.3 & -1.510 & -0.364 & 3.056\\
    3C\,351.0        &  3.2 &  22.2 &  0.0 &  0.00 &  692.4 &  8957.4 &  508.9 &  8.0 &   64.0 &  -15.6 & -1.677 & -0.961 & 3.381\\
    UGC\,10683B      &\dotfill&54.7 & 15.2 &  0.28 & 1069.2 &  4721.9 &  616.3 &\dotfill& 29.2 &   93.7 & -3.525 & -1.384 & 1.286\\
    RX\,J1715.9+3112 &\dotfill&100.5& 91.3 &  0.91 & 1376.6 &  3772.9 &  440.4 &\dotfill& 79.6 &   17.4 & -3.095 & -0.118 & 2.560\\
    IRAS\,17216+3633 & 13.8 &  43.9 & 22.6 &  0.52 &  681.0 &  4837.4 &  540.9 &  6.5 &  -50.4 &   29.2 & -1.609 & -0.964 & 1.747\\
    3C\,390.3        &  5.0 &  92.5 &  1.9 &  0.02 &  647.4 &  5070.7 &  600.8 &  8.7 &-4071.9 &  -39.3 & -2.169 &  0.393 & 2.130\\
    1H\,1934-063     &\dotfill&80.1 & 16.4 &  0.20 & 1006.0 &  5531.4 &  541.4 &\dotfill&-239.6& -226.3 & -0.859 & -0.125 & 1.861\\
    Mark\,509        &  9.5 & 146.1 & 10.4 &  0.07 &  567.4 &  3428.8 &  466.1 &  4.7 &   56.9 &  -53.4 & -0.213 & -0.759 & 2.270\\
    3C\,445.0        & 11.9 & 187.7 &  0.0 &  0.00 &  451.6 &  3269.7 &  531.6 & 15.5 & -418.0 &   16.1 & -2.379 & -0.170 & 2.245\\
    MR\,2251-178     &\dotfill&202.0&  0.0 &  0.00 & 2089.7 & 11067.5 &  938.7 &\dotfill&247.7 &   64.7 & -1.316 & -0.508 & 2.599\\
    NGC\,7469        & 11.3 &  55.3 & 14.9 &  0.27 &  784.2 &  2140.1 &  875.6 &  3.7 &   42.2 & -124.7 & -0.586 & -1.710 & 1.427\\ 
    Mark\,315        &  8.0 &  18.9 & 11.4 &  0.60 &  428.1 &  3699.5 &  725.0 &  3.8 &  -15.3 & -133.2 & -2.209 & -1.696 & 1.529\\
    Mark\,316        & 12.6 &  21.1 & 19.5 &  0.93 &  558.8 &  4689.0 &  607.6 &  4.0 &   60.4 &  -98.9 & -1.222 & -1.850 & 1.470\\  
    NGC\,7603        &\dotfill&57.5 & 43.5 &  0.76 &\dotfill&  6597.9 &  856.0 &\dotfill&\dotfill&\dotfill&-1.706& -1.443 & 1.133\\

    \enddata
    \tablenotetext{a}{In units of $\rm{km\ s^{-1}}$}
    \tablenotetext{b}{H$\beta$ broad component velocity shifts with respect to narrow component. A positive velocity indicates a 
    redshift of H$\beta_{\rm{b}}$ with respect to H$\beta_{\rm{n}}$, and a negative velocity a blueshift of H$\beta_{\rm{b}}$.}
    \tablenotetext{c}{[OIII] velocity shifts with respect to narrow component of H$\beta$. A redshift of [OIII] is indicated 
    by a positive velocity, and a blueshift by a negative velocity.}
    \tablenotetext{d}{The source shows not detectable [OIII]$\lambda$5007 emission line.}
    \end{deluxetable}

   \begin{deluxetable}{lrrrrcrrrrrrr}
   \rotate
   \tablewidth{0pt}
   \tabletypesize{\scriptsize}
   \tablecaption{Spearman Rank-Order Correlation Coefficient Matrix}
   \tablehead{
   \colhead{Property} & \colhead{EW(H$\rm{\beta_{b}}$)} & \colhead{EW(FeII)} & \colhead{RFe} & 
   \colhead{FWHM(H$\rm{\beta_{b}}$)} & \colhead{[OIII]/H$\beta_{\rm{n}}$} &
   \colhead{$\Delta \rm{V(H\beta_{b}})$} & \colhead{$\Delta \rm{V}([OIII])$} &
   \colhead{$\alpha$(60,100)} & \colhead{$\alpha$(60,25)} & \colhead{$\log(L\rm{([OIII])}$}\\
    & \colhead{(1)} & \colhead{(2)} & \colhead{(3)} & 
   \colhead{(4)} & \colhead{(5)} & \colhead{(6)} & \colhead{(7)} & \colhead{(8)} & 
   \colhead{(9)} & \colhead{(10)}\\
    } 
   
   \startdata
  (1)EW(H$\rm{\beta_{b}}$)\dotfill&...&0.272&-0.334(0.0206)&0.112&0.429(0.0111)&-0.032&0.257&0.042&0.590(0.0000)&0.522(0.003)\\
  (2)EW(FeII)\dotfill&...&...&0.692(0.000)&-0.185&-0.025&-0.087&0.159&-0.081&-0.140&-0.064\\
  (3)RFe\dotfill&...&...&...&-0.298&-0.252&0.004&-0.320(0.0319)&-0.027&-0.518(0.0003)&-0.386(0.0081)\\
  (4)FWHM(H$\rm{\beta_{b}}$)\dotfill&...&...&...&...&0.425(0.0120)&-0.018&0.343(0.0230)&-0.020&0.259&0.109\\
  (5)[OIII]/H$\beta_{\rm{n}}\dotfill$...&...&...&...&...&...&-0.331(0.050)&0.341(0.0408)&-0.006&0.538(0.0012)&0.137\\
  (6)$\Delta \rm{V(H\beta_{b}})$\dotfill&...&...&...&...&...&...&0.069&0.059&-0.039&0.124\\
  (7)$\Delta \rm{V}([OIII])$\dotfill&...&...&...&...&...&...&...&-0.286&0.248&0.006\\
  (8)$\alpha$(60,100)\dotfill&...&...&...&...&...&...&...&...&-0.210&-0.098\\
  (9)$\alpha$(60,25)\dotfill&...&...&...&...&...&...&...&...&...&0.478(0.0009)\\ 
  (10)$\log(L\rm{([OIII]))}$\dotfill&...&...&...&...&...&...&...&...&...&...\\
   \enddata
   \end{deluxetable}

  \begin{table}
  \begin{center}
  \caption{Correlations of eigenvectors with line and continuum properties}
  \begin{tabular}{ccccc}
  \tableline
  \tableline
  Property & Eigenvector 1 & Eigenvector 2 & Eigenvector 3 & Eigenvector 4\\
  \tableline
  Eigenvalue & 3.055  & 1.444  & 1.185  & 1.040 \\
  Cumulative & 33.9\% & 49.9\% & 63.2\% & 74.7\%\\
  \tableline
   EW(H$\beta_{\rm{b}}$)\dotfill     &-0.709 &-0.230  & -0.476  &  0.888\\
   RFe\dotfill                       & 0.643 & 0.302  & -0.353  &  0.162\\
   FWHM(H$\beta_{\rm{b}}$)\dotfill   &-0.537 & 0.386  &  0.108  & -0.354\\
   \lbrack OIII\rbrack/H$\beta_{\rm{n}}$\dotfill &-0.712 &  0.391 & -0.204 &  0.037\\
   $\Delta V(\rm{H}\beta_{\rm{b}})$\dotfill        & 0.174 &-0.667  &  0.380  & -0.442\\
   $\Delta V(\rm{[OIII]})$\dotfill   &-0.557 & 0.358  &  0.364  & -0.393\\
   $\alpha$(100,60)\dotfill          & 0.134 &-0.076  & -0.679  & -0.681\\
   $\alpha$(60,25)\dotfill           &-0.841 &-0.145  &  0.082  &  0.224\\
   $\log(L([\rm{OIII}]))$\dotfill         &-0.527 &-0.638  &-0.188& 0.128\\
 \tableline
  \end{tabular}
  \end{center}
  \end{table}








\begin{thebibliography}{}

  \bibitem[1997]{baker} Baker, J. C., 1997, \mnras, 286, 23
  \bibitem[1977]{bladwin} Baldwin, J. A., 1977, \apj, 214, 679
  \bibitem[2001]{barhtel} Barthel, P., 2001, \nar, 45, 591
  \bibitem[2002]{boroson} Boroson, T. A., 2002, \apj, 565, 78
  \bibitem[1992]{boroson} Boroson, T. A., \& Green, R. F., 1992, \apjs, 80, 109
  \bibitem[1994]{brotherton} Brotherton, M. S., Wills, B. J., 1994, \aaps, 184, 6003
  \bibitem[1999]{brotherton} Brotherton, M. S., van Breugel, Wil, Stanford, S. A., 1999, \apjl, 520, 87
  \bibitem[2000]{canalizo} Canalizo, G., \& Stockton, A., 2000, \apj, 528, 201 
  \bibitem[2001]{canalizo} Canalizo, G., \& Stockton, A., 2001, \apj, 555, 719
  \bibitem[1989]{cardelli} Cardelli, J. A., Clayton, G. C., \& Mathis, J. S., 1989, \aj, 345, 245
  \bibitem[1999]{chatzichristou} Chatzichristou, E. T., Vanderriest, C., \& Jaffe, W., 1999, \aap, 343, 407
  \bibitem[2001]{cid fernandes} Cid Fernandes, Heckman, T., Schmitt, H., Gonzalez Delgado, R. M., Storchi-Bergmann, T., 
          2001, \apj, 558, 81
  \bibitem[2001]{cid fernandes} Cid Fernandes, R., Storchi-Bergmann, T., \& Schmitt, H. R., 1998, \mnras, 297, 579
  \bibitem[1997]{corbin} Corbin, M., 1997a, \apjs, 113, 245
  \bibitem[1997]{corbin} Corbin, M., 1997a, \apj, 532, 136
  \bibitem[2004]{davies} Davies, R. I., Tacconi, L. J., \& Genzel, R., \apj, 602, 148
  \bibitem[1984]{de grijp} de Grijp, M. H. K., Miley, G. K., Lub, J., \& de, Jong, T., 1985, \nat, 314, 240
  \bibitem[1997]{di serego alighieri}di Serego Alighieri, S., Cimatti, A., Fosbury, R. A. E., \& Hes, R., 1997, \aap, 328, 510
  \bibitem[2003]{eracleous} Eracleous, M., \& Halpern, J. P., 2003, \apj, 599, 886
  \bibitem[1988]{evans} Evans, I. N., 1988, \apjs, 67, 373
  \bibitem[2000]{ferrarese} Ferrarese, L., \& Merritt, D., 2000, \apjl, 539, 9
  \bibitem[2000]{gebhardt} Gebhardt, K., Kormendy, J. , Ho, L. C., et al. 2000, \apj, 543, L5
  \bibitem[1999]{grupe} Grupe, D., Beuermann, K., Mannheim, K., \& Thomas, H. -C., 1999, \aap, 350, 31
  \bibitem[2004]{grupe} Grupe, D., 2004, \aj, 127, 1799
  \bibitem[2001]{gonzalez delgado} Gonzalez Delgado, R. M., Heckman, T., \& Leitherer, C., 2001, \apj, 546, 845
  \bibitem[2002]{gonzalez delgado} Gonzalez Delgado, R. M., ASPC, 528, 101
  \bibitem[1996]{granato} Granato, G. L., \& Danese, L., 1994, \mnras, 268, 235
  \bibitem[2004]{granato} Granato, G. L., De Zotti, G. Silva, L., Bressan, A., \& Danese, L., 2004, \apj, 600, 580
  \bibitem[2005]{hao} Hao, C. N., Xia, X. Y., Mao, S., Wu., Hong, \& Deng, Z. G., 2005, \apj, 625, 78 
  \bibitem[2004]{heckman} Heckman, T. M., Kauffmann, G., Brinchmann, J., et al. 2004, \apj, 613, 109
  \bibitem[1989]{jackson} Jackson, N., Browne, I. W. A., Murphy, D. W., \& Saikia, D. J., 1989, \nat, 338, 485
  \bibitem[2001]{joguet} Joguet, B., Kunth, D., Melnick, J., Terlevich, R., \& Terlevich, E., 2001, \aap, 380, 19
  \bibitem[1994]{keel} Keel, W. C., de Grijp, M. H. K., Miley, G. K., \& Zheng, W., 1994, \aap, 283, 791
  \bibitem[2001]{kewley} Kewley, L. J., Heisler, C. A., Dopita, M. A., \& Lumsden, S., \apjs, 2001, 132, 37
  \bibitem[2003]{king} King, A. R., \& Pounds, K. A., 2003, \mnras, 345, 657
  \bibitem[1996]{kriss} Kriss, G., 1996, Adass, 3, 437
  \bibitem[2000]{kuraszkiewicz} Kuraszkiewicz, J., Wilkes, B. J., Brandt, W. N., et al. 2000, \apj, 542, 631
  \bibitem[2004]{kuraszkiewicz} Kuraszkiewicz, J. K., Green, P. J., Crenshaw, D. M., Dunn, J., Forster, K., Vestergaard, M., 
          Aldcroft, T. L., 2004, \apjs, 150, 165
  \bibitem[2003]{laor} Laor, A., 2003, \apj, 590, 86
  \bibitem[1997]{laor} Laor, A., Fiore, F., Martin, E., et al. 1997, \apj, 477, 93
  \bibitem[1999]{lawrence} Lawrence, A., Elvis, M., Wilks, B. J., McHardy, I., \& Brandt, N., 1997, \mnras, 285, 879
  \bibitem[1994]{lipari} Lipari, S., 1994, \apj, 436, 102
  \bibitem[1988]{low} Low, F. J., Huchra, J. P., Kleinmann, S. G., \& Cutri, R. M., 1988, \apjl, 327, 41
  \bibitem[1998]{marorrian} Magorrian, J., Tremaine, S., Richstone, D., et al. 1998, \aj, 115, 2285
  \bibitem[2003]{marconi} Marconi, A.,; Hunt, L. K., 2003, \apjl, 589, 21
  \bibitem[2001]{marziani} Marziani, P., Sulentic, J. W.,  Dultzin-Hacyan, D., \& Calvani, M., 2001, \apj, 558, 553
  \bibitem[2003]{marziani} Marziani, P., Sulentic, J. W., Zamanov, R., \& Calvani, M., 2003a  \memsai, 74, 490
  \bibitem[2003]{marziani} Marziani, P., Zamanov, R., Sulentic, J. W., \& Calvani, M., 2003b, \mnras, 345, 1133
  \bibitem[1988]{massey} Massey, P., Strobel, K., Barnes, J. V., \& Anderson, E., 1988, \apj, 328, 315
  \bibitem[2000]{mathur} Mathur. S., 2000, \mnras, 314, L17
  \bibitem[2004]{netzer} Netzer, H., Shemmer, O., Maiolino, R., Oliva, E., Croom, S., Corbett, C., \& di Fabrizio, L., 2004, \apj, 
                614, 558  
  \bibitem[1988]{norman} Norman, C., \& Scoville, N., 1988, \apj, 332, 124
  \bibitem[1998]{o'brein} O'Brien, P. T., Dietrich, M., Leighly, K., et al. 1998, \apj, 509, 163
  \bibitem[1979]{oke} Oke, J. B., \& Lauer, T. R., 1979, \apj, 230, 360

  \bibitem[1989]{osterbrock} Osterbrock, D. E., 1989, Astrophysics of Gaseous Nebulae and Active Galactic Nuclei
                (University Science Book: Mill Valley, CA)
  \bibitem[1978]{phillips} Phillips, M. M., 1978, \apj, 226, 736
  \bibitem[1992]{pier} Pier, E. A., \& Krolik, J. H., 1992, \apj, 401, 99
  \bibitem[1998]{richstone} Richstone, D., Ajhar, E. A., Bender, R., 1998, \nat, 395, A14
  \bibitem[2000]{rodriguez-ardila} Rodriguez-Ardila, A., Pastoriza, M. G., \& Donzelli, C. J., 2000, \apjs, 126, 63
  \bibitem[2000]{rodriguez-ardila} Rodriguez-Ardila, A., \& Viegas, S. M., 2003, \apjl, 340, 33
  \bibitem[1996]{sanders} Sanders, D. B., \& Mirabel, I. F., 1996, \araa, 34, 749
  \bibitem[1988]{sanders} Sanders, D. B., Soifer, B. T., Elias, J. H., Madore, B. F., Matthews, K., Neugebauer, G., Scoville, N. Z.,
                1988, \apj, 325, 74
  \bibitem[1991]{stockton} Stockton, A., \& Ridgway, S. E., 1991, \aj, 102, 488
  \bibitem[2000]{storey} Storey, P. J., \& Zeippen, C. J., 2000, \mnras, 312, 813
  \bibitem[2003]{strateva} Strateva, I. V., Strauss, M. A., Hao, L., et al. 2003, \aj, 126, 1720
  \bibitem[2000]{sulentic} Sulentic, J. W., Marziani, P., \& Dultzin-Hacyan, D., 2000a, \araa, 38, 521
  \bibitem[2000]{sulentic} Sulentic, J. W., Zwitter, T., Marziani, P., \& Dultzin-Hacyan, D., 2000b. \apjl, 536, 5
  \bibitem[2000]{sulentic} Sulentic, J. W., Marziani, P., Zwitter, T., Dultzin-Hacyan, D., \& Calvani, 2000c, \apjl, 545, 15
  \bibitem[2002]{sulentic} Sulentic, J. W., Marziani, P., \& Zamanov, R., et al. 2002, \apjl, 566, 71
  \bibitem[2004]{sulentic} Sulentic, J. W., Stirpe, G. M., \& Marziani, P., et al. 2004, \aap, 423, 121
  \bibitem[2002]{tremaine} Tremaine, S., Gebhardt, K., Bender, R., et al. 2002, \apj, 574, 740
  \bibitem[2001]{vaughan} Vaughan, S., Edelson, R., Warwick, R. S., et al. 2001, \mnras, 327, 673
  \bibitem[1987]{veilleux} Veilleux, S., \& Osterbrock, D. E., 1987, \apjs, 63, 259
  \bibitem[2004]{veron-cetty} Veron-Cetty, M. P., Joly, M., \& Veron-Cetty, P., 2004, \aap, 417, 515
  \bibitem[2003]{veron-cetty} Veron-Cetty, M. P., \& Veron, P., 2001, \aap, 374, 92
  \bibitem[1996]{wang} Wang, T., Brinkmann, W., \& Bergeron, J., 1996, \aap, 309, 81
  \bibitem[2004]{wang} Wang, J., Wei, J. Y., \& He, X. T., 2004, ChJAA, 4, 415
  \bibitem[2005]{wang} Wang, J., Wei, J. Y., \& He, X. T., 2005, \aap, 436, 417
  \bibitem[1992]{winkler} Winkler, H., 1992, \mnras, 257, 677
  \bibitem[1997]{witt} Witt, H. J., Czerny, B., Zycki, P. T., 1997, \mnras, 286, 848
  \bibitem[2003]{xu} Xu, D. W., Komossa, S., Wei, J. Y., Qian, Y., \& Zheng, X. Z., 2003, \apj, 590, 73
  \bibitem[2002]{zamanov} Zamanov, R., \& Marziani, P., 2002, \apjl, 571, 77
  \bibitem[2002]{zamanov} Zamanov, R., Marziani, P., Sulentic, J. W., Calvani, M., Dultzin-Hacyan, D., \& Bachev, R, 2002, 
                \apjl, 576, 9
  \bibitem[2004]{zuther} Zuther, J., Eckart, A., Scharwachter, J., Krips, M., \& Straubmeier, C., 2004, \aap, 414, 919
  \bibitem[2005]{zhou} Zhou, H., Wang, T., Dong, X., Wang, J.,\& Lu, H., 2005, \memsai, 76, 93
  \bibitem[2002]{zheng} Zheng, X. Z., Xia, X. Y., Mao, S., Wu, H., \& Deng, Z. G., 2002, \aj, 124, 18



\end{thebibliography}
\end{document}